\documentclass[aps,prx,twocolumn,amsfonts,nofootinbib,preprintnumbers,superscriptaddress]{revtex4-2}
\usepackage{soul}

\usepackage{multibib}
\usepackage{url}
\usepackage{hyperref}

\usepackage[normalem]{ulem}
\usepackage{bm}
\usepackage{mwe}
\usepackage{xr}

\usepackage{amsmath,amsthm, amssymb} 
\usepackage{dsfont,yfonts}
\usepackage{hyperref}
\usepackage{floatrow,dcolumn,rotating}
\usepackage[dvipsnames]{xcolor}

\usepackage{graphics}
\usepackage{graphicx}

\hypersetup{colorlinks=true,linkcolor=blue,citecolor=red,urlcolor=blue}

\newcommand{\MeijerG}[7]{G \begin{smallmatrix} #1,\!\!&#2\\#3,\!\!&#4 \end{smallmatrix} \left( \begin{smallmatrix} #5 \\ #6 \end{smallmatrix} \middle\vert #7 \right) }

\begin{document}

\title{Multiscale network renormalization: scale-invariance without geometry}

\author{Elena Garuccio}
\affiliation{Instituut-Lorentz for Theoretical Physics, Leiden University, Niels Bohrweg 2,  2333 CA Leiden, Netherlands}

\author{Margherita Lalli}
\affiliation{IMT School for Advanced Studies, Piazza S. Francesco 19, 55100 Lucca, Italy}

\author{Diego Garlaschelli}
\affiliation{Instituut-Lorentz for Theoretical Physics, Leiden University, Niels Bohrweg 2,  2333 CA Leiden, Netherlands}
\affiliation{IMT School for Advanced Studies, Piazza S. Francesco 19, 55100 Lucca, Italy}
\affiliation{INdAM-GNAMPA Istituto Nazionale di Alta Matematica, Italy}

\begin{abstract} 
Systems with lattice geometry can be renormalized exploiting their coordinates in metric space, which naturally define the coarse-grained nodes.
By contrast, complex networks defy the usual techniques, due to their small-world character and lack of explicit geometric embedding. 
Current network renormalization approaches require strong assumptions (e.g. community structure, hyperbolicity, scale-free topology), thus remaining incompatible with generic graphs and ordinary lattices.
Here we introduce a graph renormalization scheme valid for any hierarchy of heterogeneous coarse-grainings, thereby allowing for the definition of `block-nodes' across multiple scales.
This approach identifies a class of scale-invariant networks characterized by a necessary and specific dependence on additive hidden variables attached to nodes, plus optional dyadic factors. 
If the hidden variables are annealed, they lead to realistic scale-free networks with assortativity and finite local clustering, even in the sparse regime and in absence of geometry. If they are quenched, they can guide the renormalization of real-world networks with node attributes and distance-dependence or communities. As an application, we derive an accurate multiscale model of the International Trade Network applicable across arbitrary geographic partitions.
These results highlight a deep conceptual distinction between scale-free and scale-invariant networks, and provide a geometry-free route to renormalization. 
\end{abstract}

\maketitle

\section{Introduction}
Several societal challenges, including the development of more resilient economies, the containment of infectious diseases, the security of critical infrastructures and the preservation of biodiversity, require a thorough understanding of the network structure connecting the units of the underlying complex systems~\cite{econnet,myscience,guidohumans}.
One of the obstacles systematically encountered in the analysis and modelling of real-world networks is the simultaneous presence of structures at multiple interacting scales. For instance, socioeconomic networks are organized hierarchically across several levels, from single individuals up to groups, firms, countries and whole geographical regions. Besides the interactions taking place \emph{horizontally} within the same hierarchical level (e.g. social ties among individuals or international trade relationships among countries), there are important \emph{cross-level} (e.g. individual-firm, firm-country, country-region) interactions that require a multiscale description.
Establishing a consistent representation of a graph at multiple scales is in fact a long-standing problem whose solution would enable considerable progress in the description, modelling, and control of real-world complex systems.

In the language of statistical physics, achieving a proper multiscale description of a network requires the introduction of a renormalization scheme whereby a network can be coarse-grained iteratively by partitioning nodes into `block-nodes' either horizontally, i.e. at homogeneous levels of the hierarchy, or across hierarchical levels, thus allowing block-nodes to contain possibly very different numbers of nodes.
The traditional block-renormalization scheme (whereby equally sized blocks of neighbouring nodes in a regular lattice are replaced by identical block-nodes, leading to a reduced lattice with the same geometry) is feasible for geometrically embedded networks where the coordinates of nodes naturally induce a definition of block-nodes of equal size. However, this traditional scheme becomes ill-defined in arbitrary graphs where node coordinates are not necessarily defined, and particularly problematic in real-world networks with broad degree distribution (which makes the neighbourhoods of nodes very heterogeneous in size and not good candidates as block-nodes) and small-world property (which limits the iterability of coarse-grainings based on shortest paths).
Several renormalization schemes for complex networks have been proposed to deal with these inherent complications~\cite{song2005self,gallos2007review,goh2006skeleton,laurienti2011universal,nussinov,alon,kim2004geographical,serrano2008self,krioukov2010hyperbolic,garcia2017multiscale,gfeller2007spectral,tribastone,Pablo}. For instance, in analogy with fractal analysis, a box-covering technique defining block-nodes as certain sets of neighbouring nodes has been defined~\cite{song2005self,gallos2007review,goh2006skeleton,laurienti2011universal}.
Alternative coarse-graining schemes have been proposed based on the identification of communities~\cite{nussinov} or motifs~\cite{alon}. 
Another notable approach is the geometric embedding of networks in a hidden euclidean~\cite{kim2004geographical} or hyperbolic~\cite{serrano2008self,krioukov2010hyperbolic,garcia2017multiscale} metric space, followed by the coarse-graining of nearby nodes. Hyperbolically embedded graphs have many realistic properties, including scale-free degree distributions and large clustering, that are preserved upon geometric renormalization~\cite{garcia2017multiscale}. 
Other methods are based on the preservation of certain spectral properties of the original network via the identification of (approximate) equivalence classes of structurally similar nodes~\cite{gfeller2007spectral,tribastone}. A notable recent contribution is a diffusion-based coarse-graining scheme which detects spatio-temporal scales in heterogeneous networks via the Laplacian operator for graphs~\cite{Pablo}.

Despite progress has been made, the above approaches have not yet focused on the problem of looking for the most general graph model that remains consistent across different coarse-grainings of the same network, i.e. that keeps describing the same system coherently (possibly with renormalized parameters) at all scales. 
In other words, once they find relevant aggregation levels, the available methods are in general not able to provide a random graph model of the system that remains consistent across those coarse-grainings.
Moreover, the available approaches require the existence of specific topological properties (e.g., community structure~\cite{nussinov,alon}, hyperbolicity~\cite{serrano2008self,krioukov2010hyperbolic,garcia2017multiscale}, scale-freeness~\cite{song2005self,gallos2007review,goh2006skeleton,laurienti2011universal}, approximate structural equivalence~\cite{gfeller2007spectral,tribastone},  non-trivial Laplacian susceptibility~\cite{Pablo}) and are therefore irreducible to the ordinary renormalization scheme defined for simpler lattices or (random) Euclidean graphs, which on the other hand are obvious examples of scale-invariant networks. 
Additionally, the requirement that the renormalization scheme can act flexibly across hierarchical levels in a multiscale fashion is not explicitly enforced in any of the available methods.

Here we propose a general graph renormalization scheme based on a random network model that remains invariant across all scales, for \emph{any desired (horizontal or vertical) partition of nodes into block-nodes}. In a certain `quenched' setting, the model can guide the renormalization of generic graphs, from regular lattices to realistic complex networks with node attributes and (optionally, but not necessarily) dyadic properties such as distances and/or community structure. 
In a different `annealed' setting, it can generate realistic scale-free networks spontaneously, simply as the natural result of the requirement of scale-invariance, without fine-tuning and without geometry.

The rest of the paper is organized as follows.
In Sec.~\ref{sec:model} we introduce the graph renormalization framework, identify the resulting scale-invariant network model, discuss several theoretical properties of the resulting networks, and highlight the differences with respect to the main existing models.
In Sec.~\ref{sec:quenched} we consider the quenched setting, where the model parameters are considered fixed and identifiable with empirical features, and show an application leading to a remarkably consistent one-parameter model of the International Trade Network that we validate across arbitrary geographic partitions.
In Sec.~\ref{sec:annealed} we consider the annealed setting, where the model parameters are themselves regarded as random variables subject to a scale-invariance requirement, and show how this leads spontaneously to a model of networks with interesting realistic features, including the scale-free property and a finite local clustering even in the sparse regime and in absence of any notion of metric distance.
In Sec.~\ref{sec:conclusions} we offer some concluding remarks.
Finally, in the Appendices we provide important technical details that support various results discussed in the main text. 

\section{Graph renormalization and scale-invariant network model\label{sec:model}}
\begin{figure}[t]
\includegraphics[width=\textwidth]{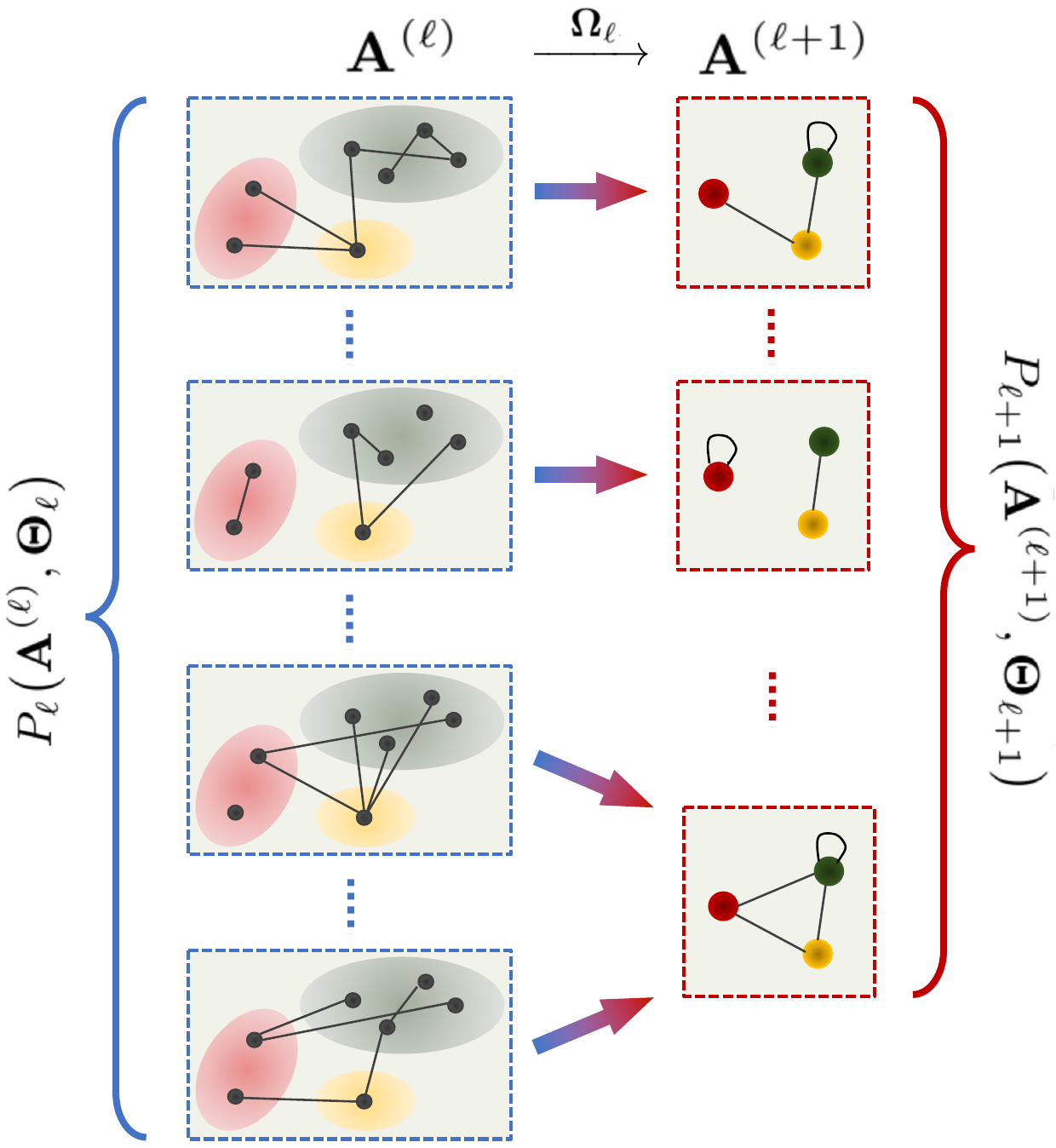}
\caption{\label{fig:coarse_graining} \textbf{Schematic example of the graph coarse-graining and induced ensembles}. Nodes of an $\ell$-graph $\mathbf{A}^{(\ell)}$ (left) are grouped together, via a given partition $\mathbf{\Omega}_\ell$, to form the block-nodes of the coarse-grained $(\ell+1)$-graph $\mathbf{A}^{(\ell+1)}$ (right). Note that, in general, block-nodes can contain different numbers of nodes.
A link between two block-nodes (or a self-loop at a single block-node) is drawn whenever a link is present between any pair of constituents nodes. 
Different realizations of the $\ell$-graph are mapped onto realizations of the $(\ell+1)$-graph via $\mathbf{\Omega}_\ell$. Multiple realizations of the $\ell$-graph may end up in the same realization of the $(\ell+1)$-graph. 
The scale-invariant requirement is obtained by viewing the realizations of the $\ell$-graph as the outcome of a random graph generating process with probability $P_\ell\big(\mathbf{A}^{(\ell)},\mathbf{\Theta}_\ell\big)$, where $\mathbf{\Theta}_\ell$ is a set of parameters, and imposing that the induced probability $P_{\ell+1}\big(\mathbf{A}^{(\ell+1)},\mathbf{\Theta}_{\ell+1}\big)$ at the next level has the same functional form as $P_\ell\big(\mathbf{A}^{(\ell)},\mathbf{\Theta}_\ell\big)$, with renormalized parameters $\mathbf{\Theta}_{\ell+1}$.}
\end{figure}

Let us consider a binary undirected graph with $N_0$ `fundamental' nodes  (labeled as $i_0=1,N_0$) and its $N_0\times N_0$ adjacency matrix $\mathbf{A}^{(0)}$ with entries $a^{(0)}_{i_0,j_0} = 1$ if the nodes $i_0$ and $j_0$ are connected, and $a^{(0)}_{i_0,j_0} = 0$ otherwise.
We do not allow for multiple edges but we do allow for self-loops, i.e. each diagonal entry can take values  $a^{(0)}_{i_0,i_0}=0,1$.
We want to aggregate the $N_0$ nodes into $N_1< N_0$ block-nodes (labeled as $i_1=1,N_1$) forming a non-overlapping partition $\mathbf{\Omega}_0$ of the original $N_0$ nodes, and connect two block-nodes if \emph{at least one link} is present between the nodes across the two blocks, as illustrated in Fig.~\ref{fig:coarse_graining}. 
Therefore the coarse-grained graph is described by the $N_1\times N_1$ adjacency matrix $\mathbf{A}^{(1)}$ with entries $a^{(1)}_{i_1,j_1} = 1-\prod_{i_0\in i_1}\prod_{j_0\in j_1}(1-a^{(0)}_{i_0,j_0})$, where $i_0\in i_1$ denotes that the chosen partition $\mathbf{\Omega}_0$ maps the original node $i_0$ onto the block-node $i_1$, i.e. $i_1=\mathbf{\Omega}_0(i_0)$. Note that we do not require $i_1\ne j_1$, as we keep allowing for self-loops as we coarse-grain (a self-loop at a block-node represents the existence of at least one link or self-loop in the subgraph connecting the original `internal' nodes). In general $i_0$ is not the only node mapped to $i_1$, i.e. $\mathbf{\Omega}_0$ is \emph{surjective}. We call $\mathbf{A}^{(0)}$ the \emph{$0$-graph} and $\mathbf{A}^{(1)}$ the \emph{$1$-graph}. 
Similarly, we call the $N_0$ nodes the \emph{$0$-nodes} and the $N_1$ block-nodes the \emph{$1$-nodes}.
Iterating the coarse-graining $\ell$ times produces a hierarchy of `blocks of blocks', with the partition $\mathbf{\Omega}_{\ell}$ leading to an $(\ell+1)$-graph with $N_{\ell+1}$ $(\ell+1)$-nodes and adjacency matrix $\mathbf{A}^{(\ell+1)}$ with entries
\begin{equation}
a^{(\ell+1)}_{i_{\ell+1},j_{\ell+1}} = 1- \prod_{i_{\ell}\in i_{\ell+1}}\prod_{j_{\ell}\in j_{\ell+1}}\left(1-a^{(\ell)}_{i_{\ell},j_{\ell}}\right)
\label{eq:Arenorm}
\end{equation}
where $i_{\ell}$ and $j_{\ell}$ are $\ell$-nodes, while $i_{\ell+1}=\mathbf{\Omega}_{\ell}(i_{\ell})$ and $j_{\ell+1}=\mathbf{\Omega}_{\ell}(j_{\ell})$ are $(\ell+1)$-nodes.

\begin{figure}[t]
\includegraphics[width=\textwidth]{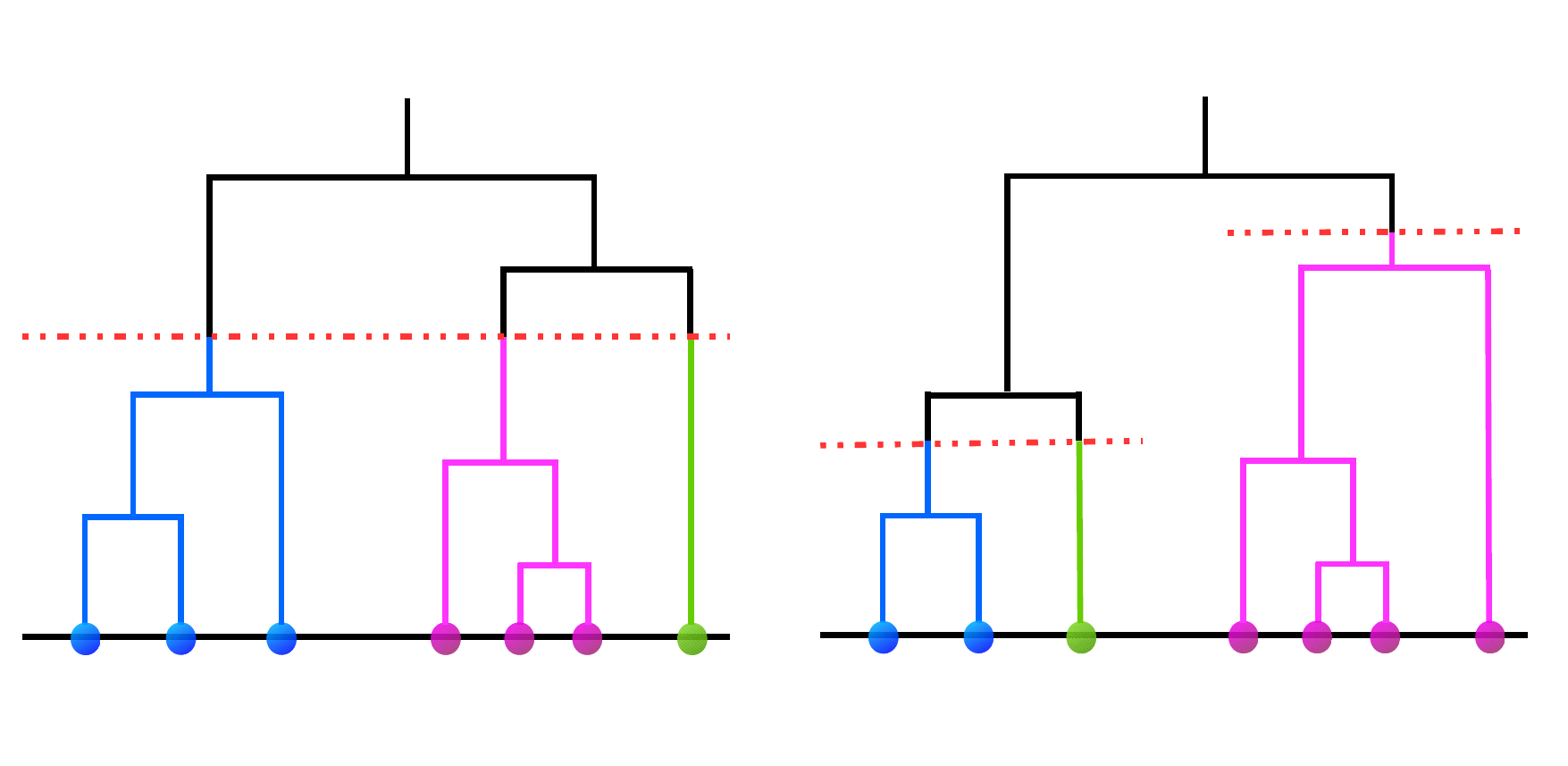}
\caption{\label{fig:dendrogram} \textbf{Horizontal vs multiscale renormalization.} Left: the desired hierarchy of coarse-grainings can be represented as a dendrogram where the $0$-nodes are the bottom `leaves' and the $\ell$-nodes are the `branches' cut out by a horizontal line placed at a suitable height. Right: if the dendrogram is cut at different heights, one obtains a multiscale renormalization scheme with block-nodes defined across multiple hierarchical levels. 
This is ultimately another partition of the $0$-nodes and is therefore readily implemented in our approach, which is designed to work for \emph{any} partition.}
\end{figure}

The hierarchy $\{\mathbf{\Omega}_\ell\}_{\ell\ge 0}$ of desired partitions can be uniquely parametrized in terms of a dendrogram as shown in Fig.~\ref{fig:dendrogram}.
Our first objective is the identification of a random graph model that can be renormalized under \emph{any} partition obtained from $\{\mathbf{\Omega}_\ell\}_{\ell\ge 0}$ via either a `horizontal' (left panel of Fig.~\ref{fig:dendrogram}) or a `multi-scale' (right panel of Fig.~\ref{fig:dendrogram}) cut of the dendrogram.
Note that, since any `multi-scale' coarse-graining is ultimately another partition of the same $0$-nodes, we can equivalently produce it `horizontally' as well, but on a certain modified hierarchy $\{\mathbf{\Omega}'_\ell\}_{\ell\ge 0}$ obtained from $\{\mathbf{\Omega}_\ell\}_{\ell\ge 0}$.
Therefore, requiring that the model is scale-invariant for any specified hierarchy of partitions automatically allows for multi-scale coarse-grainings as well.
To enforce this requirement, we fix some $\{\mathbf{\Omega}_\ell\}_{\ell\ge 0}$ and regard the initial $0$-graph $\mathbf{A}^{(0)}$ not as deterministic, but as generated by a random process with some probability $P_0\big(\mathbf{A}^{(0)},\mathbf{\Theta}_0\big)$ normalized so that $\sum_{\mathbf{A}^{(0)}\in\mathcal{G}_{N_0}}P_0\big(\mathbf{A}^{(0)},\mathbf{\Theta}_0\big)=1$, where $\mathbf{\Theta}_0$ denotes all parameters of the model (including $N_0$) and $\mathcal{G}_{N}$ denotes the set of all binary undirected graphs with $N$ nodes. 
A given partition $\mathbf{\Omega}_0$ will in general map multiple $0$-graphs $\{\mathbf{A}^{(0)}\}$ onto the same coarse-grained $1$-graph $\mathbf{A}^{(1)}$, and the notation ${\{\mathbf{A}^{(0)}\}\xrightarrow{\mathbf{\Omega}_0}\mathbf{A}^{(1)}}$ will denote such surjective mapping. Therefore $P_0\big(\mathbf{A}^{(0)},\mathbf{\Theta}_0\big)$ will induce a random process at the next level (see Fig.~\ref{fig:coarse_graining}), generating each possible $1$-graph $\mathbf{A}^{(1)}$ with probability $\sum_{\{\mathbf{A}^{(0)}\}\xrightarrow{\mathbf{\Omega}_0}\mathbf{A}^{(1)}}P_0
\big(\mathbf{A}^{(0)},\mathbf{\Theta}_0\big)$, where the sum runs over all $0$-graphs that are projected onto $\mathbf{A}^{(1)}$ by $\mathbf{\Omega}_0$.
Iterating $\ell$ times, we induce a process generating the $\ell$-graph $\mathbf{A}^{(\ell)}$ with probability $\sum_{\{\mathbf{A}^{(0)}\}\xrightarrow{\mathbf{\Omega}_{\ell-1}\cdots\mathbf{\Omega}_0}\mathbf{A}^{(\ell)}}
P_0\big(\mathbf{A}^{(0)},\mathbf{\Theta}_0\big)$, where $\mathbf{\Omega}_{\ell-1}\cdots\mathbf{\Omega}_0$ is the composition of the $\ell$ partitions $\{\mathbf{\Omega}_m\}^{\ell-1}_{m= 0}$ and ultimately represents a partition of the original $0$-nodes.

We now enforce a scale-invariant random graph model that, for any level $\ell$, can generate the possible $\ell$-graphs in two equivalent ways: either \emph{hierarchically}, i.e. by first generating the $0$-graphs with probability $P_0\big(\mathbf{A}^{(0)},\mathbf{\Theta}_0\big)$ and then coarse-graining them $\ell$ times via the partitions $\{\mathbf{\Omega}_k\}^{\ell-1}_{k= 0}$, or \emph{directly}, i.e with a certain probability $P_\ell\big(\mathbf{A}^{(\ell)},\mathbf{\Theta}_\ell\big)$ that depends on $\ell$ only through a set $\mathbf{\Theta}_\ell$ of renormalized parameters that should be obtained from $\mathbf{\Theta}_0$ using $\mathbf{\Omega}_{\ell-1}\cdots\mathbf{\Omega}_0$.
This scale-invariance requirement demands that, apart from the different dimensionality of their domains, $P_0(\cdot,\cdot)$ and $P_\ell(\cdot,\cdot)$ have the same functional form (which we denote as $P(\cdot,\cdot)$ by removing the level-dependence from the notation) and behave such that, for any pair $\ell,m$ (with $\ell> m$),
\begin{equation}
P\big(\mathbf{A}^{(\ell)},\mathbf{\Theta}_\ell\big)=
\!\!\!\!\!\!\!\!\!\!\!\!\!\sum_{\{\mathbf{A}^{(m)}\}
\xrightarrow{\mathbf{\Omega}_{\ell-1}\cdots\mathbf{\Omega}_m}\mathbf{A}^{(\ell)}}
\!\!\!\!\!\!\!\!\!\!\!\!\!\!P\big(\mathbf{A}^{(m)},\mathbf{\Theta}_m\big)\label{eq:Pml}
\end{equation}
where the renormalized parameters $\mathbf{\Theta}_\ell$ are obtained only from $\mathbf{\Theta}_m$, given $\mathbf{\Omega}_{\ell-1}\cdots\mathbf{\Omega}_{m}$. 
We look for the general solution in the case of random graphs with independent links, where $P\big(\mathbf{A}^{(\ell)},\mathbf{\Theta}_\ell\big)$ factorizes as
\begin{equation}
\prod_{i_\ell=1}^{N_\ell}\prod_{j_\ell=1}^{i_\ell}
\big[p_{i_\ell,j_\ell}\big(\mathbf{\Theta}_\ell\big)\big]^{a_{i_\ell,j_\ell}^{(\ell)}}
\big[1-p_{i_\ell,j_\ell}\big(\mathbf{\Theta}_\ell\big)\big]^{1-a_{i_\ell,j_\ell}^{(\ell)}},
\label{eq_product}
\end{equation}
where $p_{i_\ell,j_\ell}\big(\mathbf{\Theta}_\ell\big)$ is the probability that two $\ell$-nodes $i_\ell$ and $j_\ell$ are linked.
In this case it is natural to require that $\mathbf{\Theta}_\ell$ contains (besides $N_\ell$) an overall constant $\delta_\ell$ (which will set the global link density), a set of local (node-specific) `fitness' parameters $\{x_{i_\ell}\}_{i_\ell=1}^{N_\ell}$ (which will distribute the total number of links heterogeneously among nodes), and an (optional) set of dyadic (pair-specific) parameters $\{d_{i_\ell,j_\ell}\}_{i_\ell,j_\ell=1}^{N_\ell}$ (which include, when $i_\ell=j_\ell$, the `self-interaction' of a node with itself).
We can therefore use the notation 
$p_{i_\ell,j_\ell}\big(\mathbf{\Theta}_\ell\big)=p_{i_\ell,j_\ell}(\delta_\ell)$ where we keep only $\delta_\ell$ explicit in the argument of $p_{i_\ell,j_\ell}$, because the dependence on the other variables $x_{i_\ell}$, $x_{j_\ell}$, $d_{i_\ell,j_\ell}$ is already implicitly denoted by the subscripts $i_\ell$, $j_\ell$ (indeed, $p_{i_\ell,j_\ell}$ depends on $i_\ell$ and $j_\ell$ only through $x_{i_\ell}$, $x_{j_\ell}$, $d_{i_\ell,j_\ell}$).

As we show in the Appendix, if the graph probability factorizes as in Eq.~\eqref{eq_product} and the fitness $x$ is assumed to be additive upon coarse-graining of nodes, then there is a unique solution to Eq.~\eqref{eq:Pml}, given by the connection probability
\begin{equation}
p_{i_\ell,j_\ell}(\delta) = \left\{
\begin{array}{ll}
1-e^{-\delta\, x_{i_\ell}\, x_{j_\ell}\,
f(d_{i_\ell,j_\ell})}&\textrm{if}\quad i_\ell \neq j_\ell\\
1-e^{-\frac{\delta}{2}\, x^2_{i_\ell}\, f(d_{i_\ell,i_\ell})} &\textrm{if}\quad i_\ell = j_\ell\end{array}
\right.
\label{eq:prob_renorm}
\end{equation}
where $\delta>0$, $x_{i_\ell}\ge 0$ for all $i_\ell$, $f$ is an arbitrary positive function and the following renormalization rules apply:
\begin{eqnarray}
\delta_{\ell+1}&\equiv&\delta_\ell\equiv\delta,\label{eq:delta}\\
x_{i_{\ell+1}}&\equiv&\sum_{i_{\ell} \in i_{\ell+1}}x_{i_{\ell}},\label{eq:additive}\\
f\big(d_{i_{\ell+1},j_{\ell+1}}\big)&\equiv&\frac{\sum_{i_{\ell} \in i_{\ell+1}}\sum_{j_{\ell}\in j_{\ell+1}} x_{i_{\ell}} x_{j_{\ell}} f\big(d_{i_{\ell},j_{\ell}}\big)}{\sum_{i_{\ell} \in i_{\ell+1}}x_{i_{\ell}}~\sum_{j_{\ell} \in j_{\ell+1}}x_{j_{\ell}}},~\label{eq:distance}
\end{eqnarray}
i.e. $\delta$ is scale-invariant, $x$ is node-additive and $f$ renormalizes as a specific fitness-weighted average.
If the fitness is assumed to have a different renormalization rule (e.g. multiplicative rather than additive), then a corresponding modified solution is obtained (e.g. with $x$ replaced by $\log x$). So, up to a redefinition of the fitness that makes the latter additive, the solution above is general.
Note that Eq.~\eqref{eq:distance} applies also to the `diagonal' terms with $i_{\ell+1}=j_{\ell+1}$, in which case it represents the renormalized self-interaction of node $i_{\ell+1}$ with itself, determining the probability of the presence of the corresponding self-loop.

Equations~\eqref{eq:prob_renorm}-\eqref{eq:distance} are our key result.
One of their remarkable consequences is that, while the dependence of the connection probability $p_{i_\ell,j_\ell}(\delta)$ on the dyadic factor $d_{i_\ell,j_\ell}$ can be switched off entirely without destroying the scale-invariant properties of the model (e.g. by taking $f$ to be a constant function, whereby Eq.~\eqref{eq:distance} is automatically fulfilled), the dependence on the node-specific factors $x_{i_\ell}$ $x_{j_\ell}$ cannot be switched off, unless the model is made deterministic by formally requiring that $f(d_{i_\ell,j_\ell})$ takes only the two values $f=0$ (implying $p_{i_\ell,j_\ell}(\delta)=0$) or $f=+\infty$ (implying $p_{i_\ell,j_\ell}(\delta)=1$). We consider examples of both situations below.
\emph{Therefore, the dependence on dyadic factors (including geometric distances) is optional, while that on node-specific factors is necessary.}
This is a general result following only from the enforcement of scale-invariance. More specific results are discussed below.

\subsection{Scale-invariance of graph probability and partition function}
Equations~\eqref{eq:prob_renorm}-\eqref{eq:distance} have been derived using the scale-invariant requirement imposed in Eq.~\eqref{eq:Pml}, under the assumption of edge independence stated in Eq.~\eqref{eq_product}.
Indeed, inserting Eq.~\eqref{eq:prob_renorm} back into Eq.~\eqref{eq_product}, we obtain the scale-invariant graph probability explicitly as
\begin{eqnarray}
 \!\!\!&&\!\!\!\!\!\!\!\!\!P\big( \mathbf{A}^{(\ell)},\delta \big)
=\prod_{i_\ell=1}^{N_\ell}\prod_{j_\ell=1}^{i_\ell}
\!\big[1-p_{i_\ell,j_\ell}(\delta)\big]\left[\frac{p_{i_\ell,j_\ell}(\delta)}{1-p_{i_\ell,j_\ell}(\delta)}\right]^{a_{i_\ell,j_\ell}^{(\ell)}}\label{eq_product2}\\
\!\!\!&=&\displaystyle\!\prod_{i_\ell=1}^{N_\ell} \frac{
\big[e^{\frac{\delta}{2} x_{i_\ell}^2 f(d_{i_\ell,i_\ell})}\!-\!1\big]^{a_{i_\ell,i_\ell}^{(\ell)}}}{e^{\frac{\delta}{2} x_{i_\ell}^2 f(d_{i_\ell,i_\ell})}}  \prod_{j_\ell=1}^{i_\ell-1}
\!\frac{
\big[e^{\delta x_{i_\ell} x_{j_\ell} f(d_{i_\ell,j_\ell})}\!-\!1\big]^{a_{i_\ell,j_\ell}^{(\ell)}}}{e^{\delta x_{i_\ell} x_{j_\ell} f(d_{i_\ell,j_\ell})}}\nonumber\\
\!\!\!&=&\!  \frac{\displaystyle\prod_{i_\ell=1}^{N_\ell} 
\big[e^{\frac{\delta}{2} x_{i_\ell}^2 f(d_{i_\ell,i_\ell})}\!-\!1\big]^{a_{i_\ell,i_\ell}^{(\ell)}}  \prod_{j_\ell=1}^{i_\ell-1}
\!\big[e^{\delta x_{i_\ell} x_{j_\ell} f(d_{i_\ell,j_\ell})}\!-\!1\big]^{a_{i_\ell,j_\ell}^{(\ell)}}}{Q^{-1}(\delta)}\nonumber
\end{eqnarray}
where we have introduced the quantity
\begin{eqnarray}
Q(\delta) &\equiv&\prod_{i_\ell=1}^{N_\ell} e^{-\frac{\delta}{2} x_{i_\ell}^2 f(d_{i_\ell,i_\ell})} \prod_{j_\ell=1}^{i_\ell-1}
\!e^{-\delta x_{i_\ell} x_{j_\ell} f(d_{i_\ell,j_\ell})}\nonumber\\
&=&\prod_{i_\ell=1}^{N_\ell}\prod_{j_\ell=1}^{N_\ell} e^{-\frac{\delta}{2} x_{i_\ell}x_{j_\ell} f(d_{i_\ell,j_\ell})}\nonumber\\
&=&e^{-\frac{\delta}{2} \sum_{i_\ell=1}^{N_\ell}\sum_{j_\ell=1}^{N_\ell} x_{i_\ell}x_{j_\ell} f(d_{i_\ell,j_\ell})}\nonumber\\
&=&e^{-\frac{\delta}{2}x^2_{i_\infty} f(d_{i_\infty,i_\infty})}\nonumber\\
&=&1-p_{i_\infty,i_\infty}(\delta),\label{eq:Q}
\end{eqnarray}
with
\begin{eqnarray}
x_{i_\infty}&\equiv&\sum_{i_\ell=1}^{N_\ell}x_{i_\ell},\label{xinfty}\\
f(d_{i_\infty,i_\infty})&\equiv&\sum_{i_{\ell}=1}^{N_\ell}\sum_{j_{\ell}=1}^{N_\ell} \frac{x_{i_{\ell}} x_{j_{\ell}} f\big(d_{i_{\ell},j_{\ell}}\big)}{x^2_{i_\infty}}\label{finfty}
\end{eqnarray}
representing the total fitness of all nodes and the fitness-weighted average of $f$ over all pairs of nodes, respectively.
Note that our notation above suggests that $x_{i_\infty}$ and $f(d_{i_\infty,i_\infty})$ can be interpreted as the fitness and self-interaction of the supernode $i_\infty$ representing the only coarse-grained node remaining after applying an infinite sequence of partitions, or equivalently after applying the trivial partition $\Omega_\infty$ that places all nodes in the same supernode $i_\infty$ (such that $N_\infty=1$).
Indeed, when applied to such supernode, Eqs.~\eqref{eq:additive} and~\eqref{eq:distance} produce exactly the values $x_{i_\infty}$ and $f(d_{i_\infty,i_\infty})$ defined in Eqs.~\eqref{xinfty} and~\eqref{finfty}, respectively.
These quantities are obviously scale-invariant, in the sense that they can be calculated from the values taken by $x$ and $f(d)$ at any hierarchical level $\ell$. 
Therefore $Q(\delta)$ is a constant term that depends neither on the realized $\ell$-graph $\mathbf{A}^{(\ell)}$ nor, owing to Eq.~\eqref{eq:distance}, on the hierarchical level $\ell$ being considered.
Consequently, as desired, $P\big(\mathbf{A}^{(\ell)},\delta\big)$ depends on $\ell$ only through the parameters $\{x_{i_\ell}\}_{i_\ell=1}^{N_\ell}$ and $\{d_{i_\ell,j_\ell}\}_{i_\ell,j_\ell=1}^{N_\ell}$, which renormalize as stated in Eqs.~\eqref{eq:additive} and~\eqref{eq:distance}.
Note that $p_{i_\infty,i_\infty}(\delta)$ in Eq.~\eqref{eq:Q} represents the probability of a self-loop at the supernode $i_\infty$, i.e. the probability of having at least one link in the graph at any hierarchical level, so $Q(\delta)=1-p_{i_\infty,i_\infty}(\delta)$ formally represents the probability of having zero links in the network.

We can recast the above result in a way that has an explicit connection with the usual renormalization framework in statistical physics~\cite{kadanoff, wilson}. 
To do so, we rewrite the graph probability in Eq.~\eqref{eq_product2} in terms of an effective Hamiltonian $\mathcal{H}^{(\ell)}_\textrm{eff}$ and a resulting partition function $\mathcal{Z}^{(\ell)}$:
\begin{equation}
P(\mathbf{A}^{(\ell)},\delta) = \frac{e^{-\mathcal{H}^{(\ell)}_\textrm{eff}\left(\mathbf{A}^{(\ell)},\delta\right)}}{\mathcal{Z}^{(\ell)}(\delta)},\label{eq:erg}
\end{equation}
where we have defined
\begin{eqnarray}
\mathcal{H}^{(\ell)}_\textrm{eff}(\mathbf{A}^{(\ell)},\delta) &\equiv& -\sum_{i_\ell=1}^{N_\ell} \sum_{j_\ell =1}^{ i_\ell} a_{i_\ell,j_\ell} \log{\left[\frac{p_{i_\ell,j_\ell}(\delta)}{1-p_{i_\ell,j_\ell}(\delta)}\right]} \label{eq:Heff}\\
&=&-\sum_{i_\ell=1}^{N_\ell} \Big[a_{i_\ell, i_\ell}\log{ \big( e^{\frac{\delta}{2} x_{i_\ell} x_{i_\ell} f(d_{i_\ell,i_\ell}) } -1\big)}\nonumber\\
&&+\sum_{j_\ell =1}^{ i_\ell-1} a_{i_\ell, j_\ell}\log{ \big( e^{\delta x_{i_\ell} x_{j_\ell} f(d_{i_\ell,j_\ell}) } -1\big)\Big]}\nonumber
\end{eqnarray}
and
\begin{equation}
\mathcal{Z}^{(\ell)}(\delta) \equiv\sum_{\mathbf{A}^{(\ell)}} e^{-\mathcal{H}^{(\ell)}_\textrm{eff}\left(\mathbf{A}^{(\ell)},\delta\right)}.
\label{partirmorir}
\end{equation}
Note that, while Eq.~\eqref{eq:erg} is formally identical to the expression for graph probabilities in the Exponential Random Graphs (ERGs) approach~\cite{wasserman,parknewman,fronczak,mybook,mynatrevphys}, in our case it does not exhibit a \emph{sufficient statistic}, i.e. the Hamiltonian in Eq.~\eqref{eq:Heff} cannot be written as a simpler function of graph properties and, to be evaluated, requires the knowledge of the entire adjacency matrix of the graph.
A straightforward calculation yields
\begin{eqnarray}
\mathcal{Z}^{(\ell)}(\delta) & =&\sum_{\mathbf{A}^{(\ell)}} \prod_{i_\ell=1}^{N_\ell}\prod_{j_\ell=1}^{i_\ell}
\!\left[\frac{p_{i_\ell,j_\ell}(\delta)}{1-p_{i_\ell,j_\ell}(\delta)}\right]^{a_{i_\ell,j_\ell}^{(\ell)}}\nonumber\\
&=&\prod_{i_\ell = 1}^{N_\ell} \prod_{j_\ell=1}
^{ i_\ell} \sum_{a_{i_\ell,j_\ell}=0}^{1}\!\left[\frac{p_{i_\ell,j_\ell}(\delta)}{1-p_{i_\ell,j_\ell}(\delta)}\right]^{a_{i_\ell,j_\ell}^{(\ell)}}\nonumber\\
&=&\prod_{i_\ell = 1}^{N_\ell} \prod_{j_\ell=1}
^{ i_\ell} \frac{1}{1-p_{i_\ell,j_\ell}(\delta)}\nonumber\\
&=&\frac{1}{1-p_{i_\infty,i_\infty}(\delta)}\nonumber\\
&=&Q^{-1}(\delta).
\label{eq:Z}
\end{eqnarray}
As noticed above, $Q$ only depends on $\delta$, which is invariant under renormalization. We can therefore drop the superscript and denote the partition function as $\mathcal{Z}(\delta)$. Indeed, since the effective Hamiltonian $\mathcal{H}^{(\ell)}_\textrm{eff}$ has the same form given in Eq.~\eqref{eq:Heff} independently of the resolution level $\ell$, recalculating $\mathcal{Z}^{(m)}(\delta)$ from Eq.~\eqref{partirmorir} for any other coarse-graining level $m\ne \ell$ and number $N_m$ of nodes would return exactly the same value:
\begin{equation}
\mathcal{Z}^{(m)}(\delta) = \mathcal{Z}^{(\ell)}(\delta)\equiv\mathcal{Z}(\delta)=Q^{-1}(\delta),\quad\forall\, \ell,m.
\end{equation}
This means that, akin to Kadanoff's construction \cite{kadanoff}, \emph{the partition function is invariant along the renormalization flow}.
{Clearly, this property follows crucially from the functional form of the connection probability $p_{i_\ell,j_\ell}(\delta)$ in Eq.~\eqref{eq:prob_renorm} and of the induced graph probability in Eq.~\eqref{eq_product2}: any other functional form, including those considered in ERGs~\cite{wasserman,parknewman,fronczak,mybook,mynatrevphys}, would in general not lead to an invariant partition function. On the other hand, precisely because of this invariance, in our model here the effective Hamiltonian for a realization (say, $\mathbf{A}^{(m)}$) of the graph at a coarse-grained level $m$ can be evaluated exactly without knowing the microscopic details of any finer-grained version $\mathbf{A}^{(\ell)}$ (with $\ell<m$) of the same realized graph $\mathbf{A}^{(m)}$. This is not possible in ERGs, and reveals that the topology of $\mathbf{A}^{(m)}$ represents in some sense already a sort of sufficient statistic for the model, since the probability of $\mathbf{A}^{(m)}$ can be estimated without explicitly summing over the compatible topologies of any finer version of the same network, i.e. over the realizations $\{\mathbf{A}^{(\ell)}\}$ (with $\ell<m$) such that ${\mathbf{A}^{(\ell)}\xrightarrow{\mathbf{\Omega}_{m-1}\cdots\mathbf{\Omega}_\ell}\mathbf{A}^{(m)}}$ for some sequence ${\mathbf{\Omega}_{m-1}\cdots\mathbf{\Omega}_\ell}$ of partitions.}

\subsection{Node-specific fitness}
The connection probability $p_{i_\ell,j_\ell}$ increases as $x_{i_\ell}$ and/or $x_{j_\ell}$ increase. 
Therefore, as in the Fitness Model (FM)~\cite{fitness} and in the inhomogeneous random graph model (IRGM)~\cite{inho}, $x_{i_\ell}$ can be viewed as a hidden variable or `fitness' that characterizes the intrinsic tendency of the $\ell$-node $i_\ell$ to form connections. 
Here, the fitness is defined across multiple hierarchical levels and scale-invariance ensures that it is also an additive quantity summing up to the value in Eq.~\eqref{eq:additive} when $\ell$-nodes are merged onto an $(\ell+1)$-node.
This ensures that the total fitness $x_{i_\infty}$ defined in Eq.~\eqref{xinfty} is preserved by the renormalization.
For instance, if one starts with $x_{i_0}=1$ for all $i_0$, then $x_{i_\ell}$ will simply count how many $0$-nodes are found within the $\ell$-node $i_\ell$, and $x_{i_\infty}=N_0$. More interesting outcomes are obtained by using heterogeneous distributions of the fitness, as we illustrate in detail later.
We will consider both the `quenched' case where the fitness is fixed and possibly identified with some empirical quantity (thereby allowing for the renormalization of real-world networks irrespective of their scale-free behaviour), and then an opposite `annealed' scenario that spontaneously leads to scale-invariant \emph{and} scale-free networks with a density-dependent cut-off (thereby providing a generic mechanism for the emergence of scale-free networks from scale-invariance, \emph{without geometry}).

\subsection{Dyadic properties\label{sec:dyadic}}
Unlike the fitness, $d_{i_\ell,j_\ell}$ is  a dyadic factor (such as distance, similarity, co-membership in the same community, etc.) associated with the node pair $(i_\ell,j_\ell)$.
Although we are free to do otherwise, we may regard $d_{i_\ell,j_\ell}$ as a distance, in which case it may make sense to assume that $f$ is a decreasing function, ensuring that more distant nodes are less likely to be connected.
It is easy to realize that, if $d_{i_0,j_0}$ is an \emph{ultrametric distance} (i.e. such that the `stronger' triangle inequality $d_{i_0,j_0}\leq \max \left\{d_{i_0,k_0},d_{j_0,k_0}\right\}$ holds for every triple $i_0,j_0,k_0$ of $0$-nodes~\cite{ultrametricity}) and is consistent with the hierarchy of coarse-grainings (i.e. such that all distances can be represented as the heights of the branching points of the dendrogram shown in Fig.~\ref{fig:dendrogram}), then $d_{i_\ell,j_\ell}=d_{i_0,j_0}$ and hence $f(d_{i_\ell,j_\ell})=f(d_{i_0,j_0})$ whenever the $0$-nodes $i_0$ and $i_0$ map onto the $\ell$-nodes $i_\ell$ and $j_\ell$ respectively, i.e. whenever $i_\ell=\mathbf{\Omega}_{\ell-1}\cdots\mathbf{\Omega}_0(i_0)$ and $j_\ell=\mathbf{\Omega}_{\ell-1}\cdots\mathbf{\Omega}_0(j_0)$. In such a case, Eq.~\eqref{eq:distance} reduces to $f\big(d_{i_{\ell+1},j_{\ell+1}}\big)=f\big(d_{i_{\ell},j_{\ell}}\big)$ with $i_{\ell+1}=\mathbf{\Omega}_{\ell}(i_{\ell})$ and $j_{\ell+1}=\mathbf{\Omega}_{\ell}(j_{\ell})$, showing that \emph{if the distances among the $0$-nodes are ultrametric on the dendrogram induced by the hierarchy of partitions, they decouple from the hidden variables and remain invariant} across the entire coarse-graining process, just like the global parameter $\delta$. 
Reversing the point of view, we may equivalently say that, \emph{given an ultrametric distance among the $0$-nodes, any hierarchy of partitions induced by the associated dendrogram keeps the distances scale-invariant}. 
In weaker form, this also means that one may use $d_{i_0,j_0}$ to specify the dendrogram parametrizing the desired hierarchy of partitions that will keep the distances scale-invariant. 
The hierarchy may coincide with e.g. a nested community structure that one may want to impose.
In any case we stress that, although ultrametricity is an attractive property (especially in the annealed scenario that we introduce later), we do not require it as a necessary condition in general.

\subsection{Recovering the lattice case}
We can now discuss a simple but important extreme case, where the graph is constructed only as a function of distance and our approach reduces to the traditional scheme for renormalizing regular lattices. 
For instance, assume that the $0$-nodes have coordinates at the sites of a $2$-dimensional grid with lattice spacing $\tau_0$ and that $d_{i_0,j_0}$ is the Euclidean distance between these coordinates. If we set $f\equiv+\infty$ if $d_{i_\ell,j_\ell}\le 2^\ell\tau_0$ and $f\equiv0$ otherwise, then the $0$-graph will be deterministically the grid itself and the $\ell$-graph will be the usual renormalized lattice with spacing $\tau_\ell=2^\ell \tau_0$ obtained through an appropriate partition $\mathbf{\Omega}_{\ell-1}$ that maps each square block of $4$ nearest $(\ell-1)$-nodes onto a single $\ell$-node sitting at the center of the square. 
In this case, each vertical line of the dendrogram of hierarchical partitions branches regularly into $4$ `daughter' lines and $\tau_\ell=2^\ell \tau_0$ is the height of the branching points splitting $(\ell+1)$-nodes into $\ell$-nodes. 
The renormalized distances $d_{i_\ell,j_\ell}$ can be mapped exactly to this dendrogram, thereby retrieving the standard lattice renormalization scheme as a special case of our approach. Importantly, other network renormalization schemes are incompatible with this key limiting case because they require specific topologies such as scale-free degree distributions~\cite{song2005self,gallos2007review}, community structure~\cite{gfeller2007spectral,nussinov} or hyperbolic distances~\cite{serrano2008self,krioukov2010hyperbolic,garcia2017multiscale} that are obviously absent in regular grids.

\subsection{Relation to other network models}
In the opposite, more interesting extreme, the dependence on the dyadic factors can switched off. For instance, if we set $f\equiv 1$, Eq.~\eqref{eq:prob_renorm} reduces to 
\begin{equation}
p_{i_\ell,j_\ell}(\delta) = \left\{
\begin{array}{ll}
1-e^{-\delta\, x_{i_\ell}\, x_{j_\ell}}&\textrm{if}\quad i_\ell \neq j_\ell\\
1-e^{-\frac{\delta}{2}\, x^2_{i_\ell}} &\textrm{if}\quad i_\ell = j_\ell\end{array}
\right.
\label{eq:fitness}
\end{equation}
Depending on whether the fitness is considered to be quenched or annealed (a distinction that we will study in detail below), this model can also be viewed as a unique specification of the FM~\cite{fitness} or of the IRGM~\cite{inho,svante}, respectively.
{In particular, the specific form of the connection probability in Eq.~\eqref{eq:fitness} has been studied in previous works~\cite{geoff,NR,CF}. However, both our quenched (deterministic fitness) and annealed (random fitness) approaches will take a different route with respect to those previous studies. Indeed, the latter did not discuss the model in any coarse-graining setting and, importantly, considered a fitness (under the different names of `weight'~\cite{geoff}, `capacity'~\cite{NR} or `sociability'~\cite{CF}) assumed to be a random variable drawn from distributions with finite mean, whereas our fitness is either deterministic (and taken to be some fixed value measured from real data) or random but with infinite mean (the infinite-mean case being irreducible to the finite-mean one), as we shall discuss later.} 

It is important to notice that, in the `sparse' and/or `bounded' case, i.e. for $\delta\ll x_\mathrm{max}^{-2}$ and $x_\mathrm{max}<+\infty$ where $x_\mathrm{max}$ is the maximum realized (in the quenched case) or expected (in the annealed case) value of the fitness, Eq.~\eqref{eq:fitness} reduces to $p_{i_\ell,j_\ell}(\delta)\approx \delta x_{i_\ell} x_{j_\ell}$, which includes the Chung-Lu~\cite{chunglu} or `sparse' Configuration Model (CM) ($p_{i,j}\approx\delta x_i x_j$ with $x_i=k_i$ and $\delta=(2L)^{-1}$, where $k_i$ is the degree of node $i$ and $L$ is the total number of links).
Indeed, it is possible to prove the asymptotic equivalence (or a weaker form of asymptotic contiguity) of these models under certain assumptions on the expected network sparsity and on the moments of the distribution of the hidden variables~\cite{svante}.
Similarly, in the same limit Eq.~\eqref{eq:prob_renorm} reduces to $p_{i_\ell,j_\ell}(\delta)\approx \delta x_{i_\ell} x_{j_\ell}f(d_{i_\ell,j_\ell})$, which includes the sparse degree-corrected Stochastic Block-Model (dcSBM)~\cite{dcSBM} ($p_{i,j}\approx \delta x_i x_jB_{i,j}$ where $\mathbf{B}$ is a block matrix) and the Hyperbolic Model (HM)~\cite{serrano2008self,krioukov2010hyperbolic} (where $x_i$ is a `hidden degree' related to the radial coordinate of node $i$ and $d_{i,j}$ to the angular separation between nodes $i$ and $j$). The CM, dcSBM and HM are among the most popular network models and find diverse applications including community detection~\cite{fortunato}, pattern recognition~\cite{mybook} and network reconstruction~\cite{myreconstruction}.
They are examples of more general maximum-entropy random graph ensembles~\cite{mybook}, which are obtained by maximizing the entropy under constraints on certain expected structural properties~\cite{parknewman,fronczak,mynatrevphys,ginestra}. 
To generate scale-free networks with power-law degree distribution, the CM and the dcSBM are usually constructed by drawing the fitness from a power-law distribution with \emph{the same exponent}~\cite{newman_origin} of the target degree distribution (and equivalently in the HM, where the desired fitness distribution is realized via suitably sprinkling points in hyperbolic space). In the sparse regime, the fitness distribution and the degree distribution are therefore (asymptotically) identical. In the dense regime, the degree distribution has still the same power-law regime as the fitness distribution, but it additionally features a size-dependent upper cut-off, corresponding to the largest degrees approaching their maximum value~\cite{newman_origin}.

However, Eq.~\eqref{eq:fitness} is in general \emph{not} equivalent to the aforementioned models, for at least two reasons. First, in the quenched case, even if we start from a sufficiently sparse $0$-graph for which these models are consistent with Eq.~\eqref{eq:prob_renorm}, successive coarse-grainings will unavoidably increase $x_\mathrm{max}$ and bring the network to the dense regime where the CM, dcSBM and HM are described by their `full' probability $p_{i,j}=\delta x_i x_j B_{i,j}/(1+\delta x_i x_j B_{i,j})$~\cite{newman_origin,fronczak,serrano2008self,krioukov2010hyperbolic}.
Since the difference between the values of $p_{i,j}$ in Eq.~\eqref{eq:prob_renorm} and the corresponding ones in the dcSBM or HM, and similarly between those in Eq.~\eqref{eq:fitness} and the corresponding ones in the CM, are now of finite order, these models are no longer equivalent~\cite{svante} in the dense regime.
Second, in the annealed case, we will find that all moments of the distribution of the hidden variables in our approach necessarily diverge. Remarkably, this property breaks the equivalence of the different models even in the sparse case, as the conditions on the moments of the fitness distribution required for equivalence and contiguity~\cite{svante} no longer hold. 
As we show later, notable and useful consequences of this inequivalence are a non-linear dependence of the degree on the fitness (hence \emph{different} exponents of the fitness and degree distributions) and a nonvanishing local clustering coefficient even in the sparse regime.

The above considerations indicate that the multiscale model is in general \emph{not} equivalent to the CM and the dcSBM, which are not scale-invariant.
Similar considerations apply to the traditional (non-degree-corrected) SBM~\cite{SBM} (for which $p_{i,j}=B_{i,j}$) and to growing network models based on preferential attachment (PA)~\cite{PA}. In the latter, nodes enter sequentially into the network and the time at which a node enters determines its expected topological properties. 
There is no straightforward way to coarse-grain these models by defining block-nodes (possibly across different hierarchical levels) that respect the different expected properties of the nodes they contain.
The above considerations show that \emph{scale-invariant networks are consistent with a unique specification of the FM, possibly enhanced by dyadic factors, while they are incompatible with the CM, (dc)SBM and PA models.}
The connection to the Erd\H{o}s-R\'enyi (ER) model~\cite{ER} (for which $p_{i,j}=p$ for all $i,j$) is considered later in this paper, in Sec.~\ref{sec:ER}.
As for the HM, while the renormalization scheme proposed in~\cite{garcia2017multiscale} does address the consistency of the graph probability across scales, the connection probability remains congruent with the hidden metric space (i.e. retains the same functional form across coarse-grainings) only if the density of links is kept sufficiently low, such that multi-edges can be neglected. To maintain this condition enforced across multiple agglomeration levels, the HM requires a progressive pruning of links, making the scheme different from the one considered here. 
 Also, it is important to realize that, since distance-dependence has been switched off, our model in Eq.~\eqref{eq:fitness} can be renormalized exactly \emph{for any possible choice of coarse-grainings}. This shows that \emph{network renormalization does not require any notion of geometry} (whether hyperbolic or not) or spatial embedding.

\subsection{Scale-free versus scale-invariant networks} 
The above discussion sheds new light on the distinction between \emph{scale-free} networks (i.e. graphs with power-law tails in the degree distribution, as usually appearing in the CM, dcSBM and PA models) and \emph{scale-invariant} networks (i.e. graphs designed to remain consistent under agglomeration as defined here). 

The early renormalization approaches reminiscent of fractal analysis~\cite{song2005self,gallos2007review,goh2006skeleton,laurienti2011universal} relied on the idea that real-world networks can be interpreted as scale-invariant, precisely because of their scale-free property.
However the degrees, even when power-law distributed, cannot be renormalized exactly because they are neither preserved or additively transformed upon renormalization.  
The non-scale-invariance of the CM, (dc)SBM and PA models  originates precisely from the fact that their defining quantities are the node degrees. 
\emph{Unlike fractals, the self-similarity of scale-free networks applies to a topological property (the degree), not to a metric one.} 
The absence of an embedding metric space, which would provide an `ambient' dimensionality to harbour fractality in the first place (e.g. to allow for the Hausdorff-Besicovitch dimension to be strictly larger than the intrinsic topological dimension of the fractal), is also the reason why arbitrary networks cannot be easily renormalized using metric coordinates.

In general, scale-invariance as intended here is not due to the scale-free property, but to the compatibility with Eq.~\eqref{eq:prob_renorm}. 
As mentioned above, in the quenched case, and only if $\delta$ is small enough and the fitness is not too broadly distributed (so that $x_\textrm{max}<+\infty$), there may be a sparse regime where Eq.~\eqref{eq:fitness} reduces to $p_{i_\ell,j_\ell}\approx\delta x_{i_\ell},x_{j_\ell}$ with $k_{i_\ell}= x_{i_\ell}$, so that degrees are approximately additive. 
However it should be noted that, even in the latter case, degrees are rigorously additive only if each $(\ell+1)$-node is obtained as a set of $\ell$-nodes that have no mutual connection among themselves. 
This prescription is completely opposite to the more natural scheme of merging nodes that are tightly connected, e.g. because they are in the same community~\cite{nussinov} or motif~\cite{alon}.
If mutually connected nodes are mapped onto the same block-node, the degree of the latter is strictly smaller than the sum of the degrees of the original nodes. 
We may say that \emph{the coarse-graining of a network is usually designed in such a way that the additivity of degrees is maximally violated}.
In fact, this problem affects by construction all renormalization approaches based on community structure or dense motifs.
In any case, the sparse regime is destined to vanish into the dense one through the action of renormalization itself, eventually breaking the approximate additivity of degrees and producing an unavoidable upper cut-off in the degree distribution.
Moreover, we will show that in the annealed case the proportionality between fitness and degree does not hold, even in the sparse regime. In that case, scale-invariant networks have degrees that are intrinsically non-additive throughout the entire spectrum of network density.

Indeed, previous renormalization approaches based on the scale-free property encountered various problems, including lack of generality, irreducibility to the ordinary renormalization scheme in the special case of lattices, and limited iterability in small-world networks with short path lengths.
By contrast, the model proposed here can be renormalized exactly throughout the entire spectrum of network density because it is designed via a fitness that remains additive (and globally conserved at any hierarchical level) upon coarse grainings of nodes.

\section{Quenched fitness\label{sec:quenched}}
In the quenched case, the fitness of each $0$-node $i_0$ is assigned a fixed value $x_{i_0}$ and the only randomness resides in the construction of the random graph ensemble, given the fitness values. 
For instance, when modelling real-world networks, the observed nodes can be identified with the $0$-nodes and $x_{i_0}$ can be taken to be the value of some measurable additive empirical quantity attached to the $0$-node $i_0$.
Then, after choosing a hierarchy of partitions and consistently with Eq.~\eqref{eq:additive}, the fitness $x_{i_{\ell+1}}$ of each $(\ell+1)$-node $i_{\ell+1}$ (with $\ell>0$) is calculated iteratively by summing the fitness of all the $\ell$-nodes mapped onto $i_{\ell+1}$.
For each pair $(i_0,j_0)$ of $0$-nodes, a distance $d_{i_0,j_0}$ may also be specified (and possibly measured from empirical data as well) and used to determine $f(d_{i_0,j_0})$. 
Consistently, the quantity $f(d_{i_{\ell+1},j_{\ell+1}})$ between each pair $(i_{\ell+1},j_{\ell+1})$ of $(\ell+1)$-nodes is calculated via Eq.~\eqref{eq:distance}.
Together, fitness and distance determine the probability~\eqref{eq:prob_renorm} of connection between nodes at all scales.
Clearly, once $f$ is specified, the only free parameter is $\delta$, controlling the overall density of the random network.
When considering real-world networks for which fitness and distance can be measured from empirical data separately from the network structure, we may use the quenched model in order to check whether Eq.~\eqref{eq:prob_renorm} reproduces the observed topological properties of the $0$-graph itself and, if this is the case, to provide a testable multi-scale model of the renormalized network at any higher level of aggregation.

\subsection{The International Trade Network}
To illustrate this procedure, we consider the empirical International Trade Network (ITN), using the BACI-Comtrade dataset \cite{gaulier2010baci} which reports the international trade flows (imports and exports) between all pairs of world countries. We show the results for the year 2011; we have obtained similar results for the other years available in the database. 
We select this particular network because previous research has clarified that the topology of the ITN is strongly dictated by the GDP of countries~\cite{mywtw,mydouble,myassaf1,myassaf2}. Moreover, the economics literature has extensively shown that both GDP and geographical distance are key determinants of international trade, leading to the so-called `Gravity Model' of trade~\cite{gravity,fagiologravity}.
The additivity of the GDP (i.e. the aggregate GDP of two countries is the sum of their GDPs) makes the ITN a perfect candidate for our analysis, and allows us to introduce a novel renormalization scheme for this important economic network across arbitrary levels of geographical aggregation. In particular, our aim is twofold. On the one hand, we want to introduce a multiscale model of the ITN derived from first principles, i.e. using the unique combination of GDP and geographical distances dictated by Eq.~\eqref{eq:prob_renorm}, rather than arbitrary or data-driven combinations. 
On the other hand, we want to check whether the empirical topology of the ITN is consistent with the multiscale model not only at the country level at which it is usually studied (here, the $0$-graph), but also across different hiearchical levels using the renormalization rules in Eqs.~\eqref{eq:additive} and~\eqref{eq:distance}.

First, we define the multiscale model of the ITN.
We identify each $0$-node $i_0$ with a specific country for which there are GDP data available from the World Bank~\cite{worldbank} in the considered year. This results in $N_0=183$ $0$-nodes (see the Appendix).
Then, we set the fitness $x_{i_0}$ of each $0$-node equal to the empirical value of the GDP: $x_{i_0}=\mathrm{GDP}_{i_0}$, $i_0=1,N_0$. 
For each pair $(i_0,j_0)$ of countries, we also set the distance $d_{i_0,j_0}$ equal to the empirical geographical distance between the corresponding countries, using the BACI-CEPII GeoDist~\cite{mayer2011notes,thetadist} database that reports population-averaged inter-country distances (see the Appendix).
Next, we use these distances to induce a hierarchy of partitions $\{\mathbf{\Omega}_\ell\}_{\ell\ge 0}$ that define the possible coarse-grainings of the ITN. 
Technically, this is done by merging geographically close countries into `block-countries' following a single-linkage hierarchical clustering algorithm based on the GeoDist distances $\{d_{i_0,j_0}\}_{i_0,j_0=1}^{N_0}$. The output of this algorithm is a dendrogram (shown in the Appendix) like the one in Fig.~\ref{fig:dendrogram}, where the leaves are the original countries ($0$-nodes), the branching points are the block-countries, and the height of each branching point represents the ultrametric geographical distance between pairs of countries across the corresponding two branches (the ultrametric distances $\{d^<_{i_0,j_0}\}_{i_0,j_0=1}^{N_0}$ obtained via the single-linkage clustering is known as \emph{subdominant ultrametric} distances and ensure the smallest possible distortion among all possible ultrametric distances approximating the original metric distances `from below'~\cite{ultrametricity}). Cutting the dendrogram at a fixed height $h_\ell$ defines the hierarchical level $\ell$ and identifies a unique partition $\mathbf{\Omega}_\ell$ of countries into a certain number $N_\ell$ of `$\ell$-countries'. This partition can be regarded as a multiscale aggregation of countries into groups of varying size, following from actual geographical closeness rather than pre-imposed regional or political criteria. Cutting the dendogram at multiple heights $\{h_\ell\}_{\ell\ge 0}$ (with $h_0=0$) identifies a set $\{\ell\}$ of hierarchical levels, a geography-induced hierarchy $\{\mathbf{\Omega}_\ell\}_{\ell\ge 0}$ of partitions, and a corresponding sequence $\{N_\ell\}_{\ell\ge 0}$ of numbers of block-countries.
We considered 18 hierarchical levels (from ${\ell=0}$ to $\ell=17$), such that the number of block-countries is $N_\ell=183$ for $\ell=0$ and $N_\ell=180-10\ell$ for $\ell=1,17$. 
For each of these levels, the additivity of GDP ensures that Eq.~\eqref{eq:additive} holds as a definition for the empirical aggregate GDP of block-countries:
\begin{equation}
\mathrm{GDP}_{i_{\ell+1}}\equiv\sum_{i_{\ell} \in i_{\ell+1}}\mathrm{GDP}_{i_{\ell}}.
\label{eq:GDP}
\end{equation}
We then fix the function $f$ in Eq.~\eqref{eq:prob_renorm}
as $f(d)=d^{-1}$, so that the renormalized geographical distances equal
\begin{equation}
d^{-1}_{i_{\ell+1},j_{\ell+1}}\equiv\frac{\sum_{i_{\ell} \in i_{\ell+1}}\sum_{j_{\ell}\in j_{\ell+1}} \!\mathrm{GDP}_{i_{\ell}}\, \mathrm{GDP}_{j_{\ell}}\, d_{i_{\ell},j_{\ell}}^{-1}}{\sum_{i_{\ell} \in i_{\ell+1}}\!\mathrm{GDP}_{i_{\ell}}~\sum_{j_{\ell} \in j_{\ell+1}}\!\mathrm{GDP}_{j_{\ell}}},
\label{eq:geodist}
\end{equation}
which is the GDP-averaged equivalent of the population-averaged distances commonly used in geography and in the GeoDist database itself~\cite{mayer2011notes} (see the Appendix).
In this way, $d_{i_{\ell+1},j_{\ell+1}}$ represents a sort of distance between the `barycenters' of the block-countries $i_{\ell+1}$ and $j_{\ell+1}$, where the barycenter of each $(\ell+1)$-country is defined via the internal GDP distribution across the constituent $\ell$-countries. 
Note that, if we used the subdominant ultrametric distances $\{d^<_{i_0,j_0}\}_{i_0,j_0=1}^{N_0}$ produced by the single-linkage clustering algorithm rather than the original distances $\{d_{i_0,j_0}\}_{i_0,j_0=1}^{N_0}$, then Eq.~\eqref{eq:geodist} would reduce to $d^<_{i_{\ell+1},j_{\ell+1}}=d^<_{i_{\ell},j_{\ell}}$ (no effect of renormalization on ultrametric distances) as discussed in Sec.~\ref{sec:dyadic}.  
However, to remain coherent with the GeoDist averaging approach, we use the original distances  $\{d_{i_0,j_0}\}_{i_0,j_0=1}^{N_0}$.
Putting all the above ingredients together, we arrive at the following multiscale model for the ITN:
\begin{equation}
p_{i_\ell,j_\ell}(\delta) = \begin{cases}
 1- e^{-\delta\, \mathrm{GDP}_{i_\ell} \mathrm{GDP}_{j_\ell} /d_{i_\ell,j_\ell}} &\textrm{if}\quad i_\ell \neq j_\ell\\
  1- e^{-\frac{\delta}{2}\,\mathrm{GDP}^2_{i_\ell}/d_{i_\ell,i_\ell}} & \textrm{if}\quad i_\ell = j_\ell
\end{cases}
\label{eq:prob_wtw}
\end{equation}
where $\delta$ is the only free parameter and where the renormalization rules are given by Eqs.~\eqref{eq:GDP} and~\eqref{eq:geodist}.

Now that we have defined our multiscale model of the ITN, we build the corresponding instances of the real network at the chosen 18 levels of aggregation.
To this end, we construct the empirical $0$-graph $\tilde{\mathbf{A}}^{(0)}$ by drawing an undirected link between each pair of countries that have a positive trade relationship in either direction in the BACI-Comtrade dataset (see the Appendix). 
Then, we use the distance-induced partitions $\{\mathbf{\Omega}_\ell\}_{\ell\ge 0}$ defined above in order to construct the $\ell$-graph according to Eq.~\eqref{eq:Arenorm} for each level $\ell$.
This procedure creates a sequence $\{\tilde{\mathbf{A}}^{(\ell)}\}_{\ell\ge0}$ of empirical coarse-grained versions of the ITN, each one representing the existence of trade among $\ell$-countries.

\begin{figure*}[t]
\includegraphics[width=\textwidth]{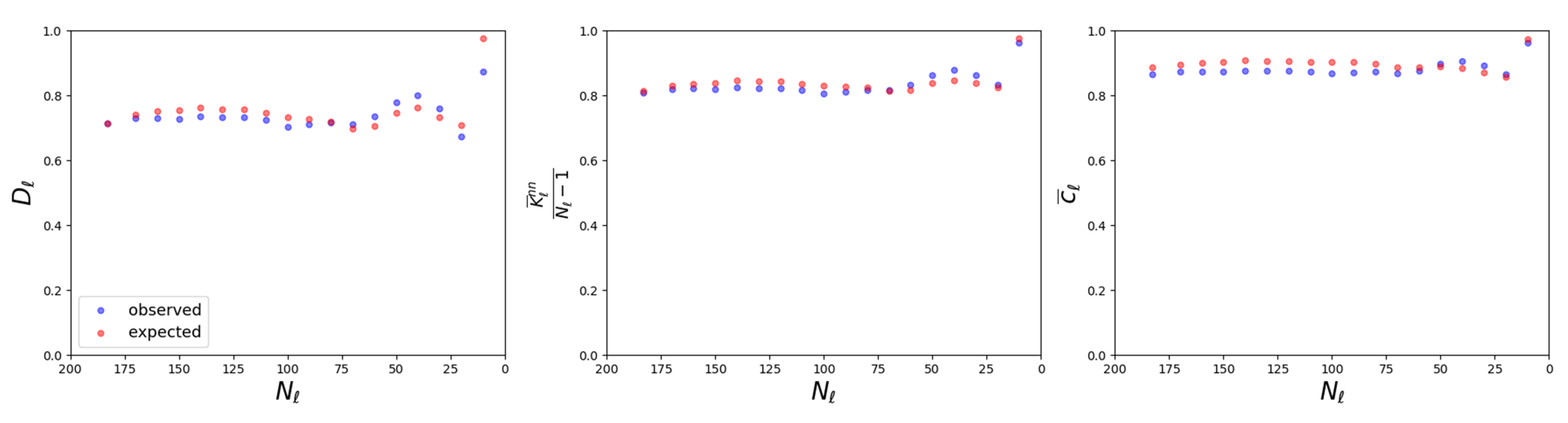}
\caption{\label{fig:global} \textbf{Prediction of global topological properties of the renormalized ITN across the full spectrum of geographical aggregation using the multiscale model.}
The panels show the agreement between the empirical and expected values of the link density $D_\ell$ including possible self-loops (left), node-averaged rescaled average nearest neighbour degree $\bar{k}^{nn}_\ell/(N_\ell-1)$ (middle) and node-averaged local clustering coefficient $\bar{c}_\ell$ (right) as functions of the number $N_\ell$ of $\ell$-countries, for all the 18 hierarchical levels considered ($\ell=0,17$).}
\end{figure*}

We can now test the multiscale model defined by Eq.~\eqref{eq:prob_wtw} against the real data $\{\tilde{\mathbf{A}}^{(\ell)}\}_{\ell\ge0}$.
Preliminarily, we calibrate the model by setting $\delta$ to the unique value $\tilde{\delta}$ that produces the same link density $D_0$ as the real ITN, i.e. such that the expected number of links in the $0$-graph (that is a monotonically increasing function of $\delta$) equals the empirical value observed in $\tilde{\mathbf{A}}^{(0)}$ (see the Appendix).
After this single parameter choice, all the probabilities in Eq.~\eqref{eq:prob_wtw} are uniquely determined at all hierarchial levels and we can test the model by comparing the empirical and expected value of various topological properties of the ITN at different coarse-grainings. 
In particular, for each level $\ell$ we consider the \emph{link density} $D_\ell$ (including possible self-loops) and, for each $\ell$-node $i_\ell$, the \emph{degree} $k_{i_\ell}$, the \emph{average nearest neighbour degree} $k^{nn}_{i_\ell}$~\cite{newman_origin} and the \emph{local clustering coefficient} $c_{i_\ell}$~\cite{local_clust} (see the Appendix for all definitions).

As a first \emph{global} test of the model, in Fig.~\ref{fig:global} we plot, for each hierarchical level ($\ell=0,17$), the link density $D_\ell$, the normalized overall average nearest neighbour degree $\bar{k}^{nn}_{\ell}/(N_\ell-1)$ and the overall local clustering coefficient $\bar{c}_{\ell}$  as a function of the number $N_\ell$ of $\ell$-nodes (the bar over a quantity denoting an average over all $\ell$-nodes). 
Note that all these global quantities are normalized on the same interval $[0,1]$, irrespective of $\ell$.
We see that the model remains in accordance with the empirical values for a wide range of hierarchical levels.
This is remarkable, given that the model has only one free parameter ($\delta$), which was calibrated uniquely to match the density $D_0$ of the $0$-graph, while the agreement holds for the other quantities as well, and across multiple levels.
This consistency across scales is an evidence of the desirable property of \emph{projectivity}~\cite{pim}.
Interestingly, all the rescaled quantities remain roughly constant as the level increases (i.e. as $N_\ell$ decreases). In line with our previous discussion about the non-equivalence between Eq.~\eqref{eq:prob_renorm} and the CM and dcSBM, the large values of density confirm that our model is necessarily different from the model that would be obtained by inserting the GDP into the equations for the CM or dcSBM.

\begin{figure*}[t]
\includegraphics[width=.8\textwidth]{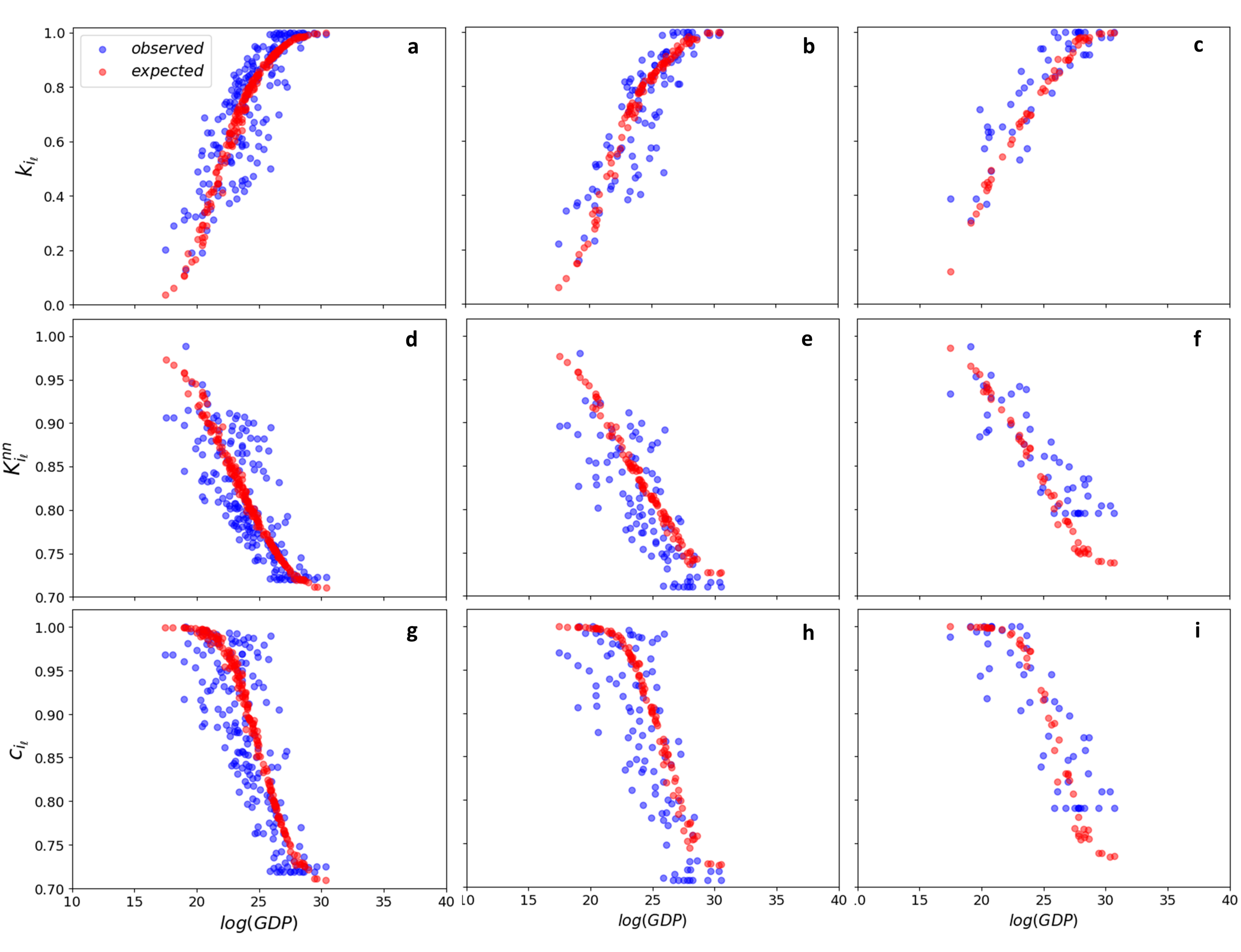}
\caption{\label{fig:node} \textbf{Prediction of local topological properties of the renormalized ITN across the full spectrum of geographical aggregation using the multiscale model.} Top panels (a,b,c): empirical (blue) and expected (red) degree $k_{i_\ell}$ vs $\ln(\mathrm{GDP}_{i_\ell})$ for all $N_\ell$ nodes, for three representative hierarchical levels ($\ell_1=0$, $\ell_2=8$, $\ell_3=13$) such that $N_{\ell_1}=183$ (left), $N_{\ell_2}=100$ (centre) and $N_{\ell_3}=50$ (right). Middle panels (d,e,f): empirical (blue) and expected (red) average nearest-neighbour degree $k^{nn}_{i_\ell}$ vs $\ln(\mathrm{GDP}_{i_\ell})$ for all $N_\ell$ nodes, for the same three hierarchical levels. Bottom panels (g,h,i): empirical (blue) and expected (red) local clustering coefficient $c_{i_\ell}$ vs $\ln(\mathrm{GDP}_{i_\ell})$ for all $N_\ell$ nodes, for the same three hierarchical levels. }
\end{figure*}

As an even more stringent test of the model, in Fig.~\ref{fig:node} we confirm the prediction that the \emph{local} topological properties of the individual (block-)countries, and in particular $k_{i_\ell}$, $k^{nn}_{i_\ell}$ and $c_{i_\ell}$, should depend strongly on the empirical value of $\mathrm{GDP}_{i_\ell}$, in a way that is governed by Eq.~\eqref{eq:prob_wtw} at all levels.
As shown in the figure, the model predictions are confirmed by the empirical data. 
It is remarkable that the agreement between observations and model expectations holds locally at the level of individual nodes and across all hierarchical levels, despite the fact that, as already noted, the single parameter $\delta$ was used to match only the density of the $0$-graph, which is a global property defined at a single hierarchical level.
As a final consistency check, and a further evidence of projectivity, we also confirmed that results similar to those shown in Figs.~\ref{fig:global} and~\ref{fig:node} are retrieved if $\delta$ is initially fixed in order to match the empirical density of $\tilde{\mathbf{A}}^{(\ell)}$ for any other given level $\ell>0$ (not shown).

All the above results confirm that there is a profound difference between scale-invariant and scale-free networks: the ITN is definitely not a scale-free network (its degree distribution is not power-law~\cite{mywtw,mydouble,myassaf1,myassaf2}, and in any case could be turned into virtually any distribution via an \emph{ad hoc} coarse-graining), yet its structure turns out to be remarkably scale-invariant.

\section{Annealed fitness\label{sec:annealed}}
In the annealed case we regard not only the graph structure, but also the fitness as a random variable. 
At the $0$-th level, this means that, for all $i_0=1,N_0$, the value $x_{i_0}$ is drawn from  from a certain probability density function (PDF) $\rho_{i_0}(x,\mathbf{\Gamma}_{i_0})$ with positive support, where $\mathbf{\Gamma}_{i_0}$ denotes all parameters of the PDF.
As for the randomness in the topology, we impose that the randomness in the fitness, induced from $\{x_{i_0}\}_{i_0=1}^{N_0}$ to  $\{x_{i_\ell}\}_{i_\ell=1}^{N_\ell}$ at all higher levels $\ell>0$ by the additivity property in Eq.~\eqref{eq:additive}, should be scale-invariant.
This means that we should be able to produce the possible values of $x_{i_\ell}$ with exactly the same probability by proceeding along two equivalent ways: hierarchically by sampling each value $x_{i_0}$ from its PDF $\rho_{i_0}(x,\mathbf{\Gamma}_{i_0})$ and summing up these values for all the $0$-nodes that are mapped onto $i_\ell$ by the partition $\mathbf{\Omega}_{\ell-1}\cdots\mathbf{\Omega}_0$, or directly by drawing $x_{i_\ell}$ from a certain PDF $\rho_{i_\ell}(x,\mathbf{\Gamma}_{i_\ell})$ that should have the same functional form of $\rho_{i_0}(x,\mathbf{\Gamma}_{i_0})$ and a set of renormalized parameters $\mathbf{\Gamma}_{i_\ell}$ obtainable from $\{\mathbf{\Gamma}_{i_0}\}_{i_0=1}^{N_0}$ only through the kwnoledge of $\mathbf{\Omega}_{\ell-1}\cdots\mathbf{\Omega}_0$. 
In other words, the fitness values can be virtually \emph{resampled} at each scale $\ell$ from a universal distribution with scale-invariant functional form and possibly scale-dependent parameters.

The above requirement is equivalent to imposing that $\rho_{i_\ell}(x,\mathbf{\Gamma}_{i_\ell})$ belongs to the family of $\alpha$-\emph{stable distributions}~\cite{levy1925calcul},  which are characterized by the four parameters $\mathbf{\Gamma}_{i_\ell}\equiv(\alpha_{i_\ell},\beta_{i_\ell},\gamma_{i_\ell},\mu_{i_\ell})$ where $\beta_{i_\ell} \in [-1,1]$, $\mu_{i_\ell} \in  \mathbb{R}$ and $\gamma_{i_\ell} > 0$ control the skewness, location and scale of $\rho_{i_\ell}(x,\mathbf{\Gamma}_{i_\ell})$ respectively, while $\alpha_{i_\ell}\equiv\alpha\in(0,2]$ is the (invariant) stability parameter, equal to the exponent asymptotically characterizing (if $\alpha<2$) the power-law tails of the distribution, i.e. $\rho_{i_\ell}(x,\mathbf{\Gamma}_{i_\ell})\propto |x|^{-1-\alpha}$ for $x$ large. For $\alpha=2$, $\rho_{i_\ell}(x,\mathbf{\Gamma}_{i_\ell})$ is instead Gaussian.
The Gaussian ($\alpha=2$), Cauchy  ($\alpha=1$, $\beta_{i_\ell}=0$) and L\'evy ($\alpha=1/2$, $\beta_{i_\ell}=1$) distributions are the only $\alpha$-stable distributions known in closed form. 
Despite this limitation, the \emph{characteristic function} (CF) 
\begin{equation}
\varphi_{i_\ell}(t,\mathbf{\Gamma}_{i_\ell})\equiv\int_{-\infty}^{+\infty} e^{itx}\,\rho_{i_\ell}(x,\mathbf{\Gamma}_{i_\ell})\,\mathrm{d}x
\end{equation}
of a general $\alpha$-stable distribution is completely known~\cite{levy1925calcul}:
\begin{equation}
\varphi_{i_\ell}(t,\mathbf{\Gamma}_{i_\ell})=\left\{\begin{array}{ll} e^{i t \mu_{i_\ell}-|\gamma_{i_\ell} t|^{\alpha}\big[1-i \beta_{i_\ell} \rm{sign}(t) \tan\frac{\pi \alpha}{2}\big]}&\mbox{if }\alpha \ne 1,\\ 
e^{i t \mu_{i_\ell}-|\gamma_{i_\ell} t | \big[1+i \beta_{i_\ell} \frac{2}{\pi}\rm{sign}(t)\ln|t|\big]}&\mbox{if }\alpha = 1.
\end{array}\right.\nonumber
\end{equation}

A key feature of $\alpha$-stable distributions is that, under the additive rule stated in Eq.~\eqref{eq:additive}, the parameters renormalize as
\begin{eqnarray}
\alpha_{i_{\ell+1}}&\equiv&\alpha,\label{eq:alpha}\\
\beta_{i_{\ell+1}}&\equiv&\frac{\sum_{i_{\ell} \in i_{\ell+1}} \beta_{i_{\ell}} \gamma_{i_{\ell}}^{\alpha}}{\sum_{i_{\ell} \in i_{\ell+1}}  \gamma_{i_{\ell}}^{\alpha}},\label{eq:beta}\\ 
\gamma_{i_{\ell+1}}^{\alpha}& \equiv &\sum_{i_{\ell} \in i_{\ell+1}}  \gamma_{i_{\ell}}^{\alpha},\label{eq:gamma}\\
\mu_{i_{\ell+1}}&\equiv&\sum_{i_{\ell} \in i_{\ell+1}}  \mu_{i_{\ell}}.\label{eq:mu} \end{eqnarray}
When $0<\alpha<1$ and $\beta_{i_\ell}=1$, the support of $\alpha$-stable distributions is $[\mu_{i_\ell},+\infty)$. In order to ensure non-negative fitness values at all scales $\ell\ge0$ (as required in the connection probability $p_{i_\ell,j_\ell}$), we therefore start from $\ell=0$ and set $0<\alpha<1$ and $\beta_{i_0}=1$, $\mu_{i_0}=0$ for all $i_0=1,N_0$. Note that we might set $\mu_{i_0}>0$ as well, but in that case Eq.~\eqref{eq:mu} would imply an increase of $\mu_{i_\ell}$ with $\ell$, while we do not want to progressively restrict the possible values of the fitness as $\ell$ increases; in other words, we want to keep the support of the fitness distribution scale-invariant. 
With this choice, Eqs.~\eqref{eq:alpha}-\eqref{eq:mu} imply that, at all higher levels,
\begin{equation}
\alpha_{i_{\ell+1}}\!\equiv\alpha\in(0,1),\, \beta_{i_{\ell+1}}\!\equiv1,\, \gamma_{i_{\ell+1}}^{\alpha}\!\equiv\!\!\sum_{i_{\ell} \in i_{\ell+1}}  \gamma_{i_{\ell}}^{\alpha},\,
\mu_{i_{\ell+1}}\!\equiv0,\nonumber
\end{equation}
showing that $\alpha$, $\beta$ and $\mu$ are scale-invariant, while $\gamma^{\alpha}$ is node-additive.

The above scaling rules, combined with the form of $\varphi_{i_\ell}(t,\mathbf{\Gamma}_{i_\ell})$ given above, finally lead to the scale-invariant CF of the fitness, for all $\alpha\in(0,1)$ and for all $\gamma_{i_\ell}>0$:
\begin{equation}
\varphi_{i_\ell}(t,\alpha,\gamma_{i_\ell})= e^{-|\gamma_{i_\ell} t|^{\alpha}\big[1-i\,\rm{sign}(t) \tan\frac{\pi \alpha}{2}\big]}.
\label{eq:stable}
\end{equation}
This choice corresponds to the so-called class of \emph{one-sided stable distributions}~\cite{onesided1,onesided2,onesided3,onesided4,onesided5}.
For this particular class it is also known that, up to a scale transformation reabsorbed in the value of $\gamma_\alpha\equiv[\textrm{cos}(\alpha\pi/2)]^{1/\alpha}$, the Laplace transform (LT) $\lambda_{i_\ell}(t,\alpha,\gamma_\alpha)$ of the PDF $\rho_{i_\ell}(x,\alpha,\gamma_\alpha)$ equals
\begin{equation}
\lambda_{i_\ell}(t,\alpha,\gamma_\alpha)\equiv\int_0^{+\infty}  e^{-tx}\rho_{i_\ell}(x,\alpha,\gamma_\alpha)\mathrm{d}x=e^{{-t}^\alpha}\label{eq:laplace}
\end{equation}
which is a stretched exponential~\cite{onesided1,onesided2,onesided3,onesided4,onesided5}.
Importantly, the requirement $\alpha\in(0,1)$ implies that \emph{all moments of the fitness distribution diverge} (including the mean). As anticipated above, this in turn implies that the models in Eqs.~\eqref{eq:prob_renorm} and~\eqref{eq:fitness}, even in the sparse case, are \emph{not} equivalent to the dcSBM and CM respectively, as the conditions for equivalence~\cite{svante} break down.
This shows that the annealed versions of the dcSBM and the CM are not scale-invariant, even in the sparse regime.
In order to work with an explicit scale-invariant PDF of the fitness, we can use the only stable distribution known in closed form within the above constraints, i.e. the L\'evy distribution for which $\alpha = 1/2$:
\begin{equation}
\rho_{i_\ell}(x,1/2,\gamma_{i_\ell}) = \sqrt{\frac{\gamma_{i_\ell}}{2\pi} }\frac{e^{-\gamma_{i_\ell}/(2x)}}{x^{3/2}}\quad (x>0),
\label{eq:Levy}
\end{equation} 
where we have restored the arbitrary parameter $\gamma_{i_\ell}>0$, which is the only remaining free parameter and is subject to renormalization rule given by Eq.~\eqref{eq:gamma}. 

In summary, in the annealed scenario, at any hierarchical level $\ell$ the fitness of each $\ell$-node is a random variable described by the CF $\varphi_{i_\ell}(t,\alpha,\gamma_{i_\ell})$ in Eq.~\eqref{eq:stable} or equivalently by the LT $\lambda_{i_\ell}(t,\alpha,\gamma_\alpha)$ in Eq.~\eqref{eq:laplace}. If $\alpha = 1/2$, the PDF is known explicitly from Eq.~\eqref{eq:Levy} and such that $\rho_{i_\ell}(x,1/2,\gamma_{i_\ell})\propto x^{-3/2}$ for $x$ large, while for general $\alpha\in(0,1)$ we know that $\rho_{i_\ell}(x,\alpha,\gamma_{i_\ell})\propto x^{-1-\alpha}$ for $x$ large, even if the explicit form is not known. Given a realization of these fitness values, the network is generated with probability $P\big(\mathbf{A}^{(\ell)},\delta\big)$ given by Eq.~\eqref{eq_product2}, i.e. by connecting pairs of $\ell$-nodes with connection probability $p_{i_\ell,j_\ell}(\delta)$ given by Eq.~\eqref{eq:prob_renorm}.
This construction is entirely self-consistent across all hierarchical levels, i.e. the $\ell$-graph can be either be built bottom-up, starting from level $0$ and coarse-graining the $0$-graph up to level $\ell$, or directly at the $\ell$-th level, by sampling the fitness at that level and generating the resulting $\ell$-graph immediately.
Note that, up to this point, the connection probability $p_{i_\ell,j_\ell}$ can still depend on the distances $d_{i_\ell,j_\ell}$ as long as the latter are ultrametric on the histogram of desired coarse grainings and therefore decoupled from the fitness, as discussed previously (if the distances between $0$-nodes are not ultrametric, Eq.~\eqref{eq:distance} would make the renormalized distances fitness-dependent and hence random in the annealed case).

In the rest of this section, we provide a series of analytical results for the case $\alpha=1/2$ (which corresponds to the only stable distribution known in closed form in the range of interest for $\alpha$) and various numerical results for other values of $\alpha\in(0,1)$. In a companion paper~\cite{rajat}, we provide more rigorous mathematical proofs for all values of $\alpha\in(0,1)$ by replacing the $\alpha$-stable distribution for the fitness with a Pareto distribution with the same tail exponent $-1-\alpha$, in order to make the problem more analytically tractable. Whenever relevant, we refer to those results in what follows.

\subsection{From semi-group to group}
Notably, a unique property of the annealed case is that the renormalization procedure defines not only a \emph{semi-group} proceeding bottom-up from the $0$-graph to higher levels as in usual schemes, but also a \emph{group}: it can proceed top-down as well, by resolving the $0$-graph into a graph with any number of nodes bigger than $N_0$, indefinitely and in a scale-invariant manner. This possibility is ensured by the fact that stable distributions are \emph{infinitely divisible}, i.e. they can be expressed as the probability distribution of the sum of an arbitrary number of i.id. random variables from the same family.
This property implies that we can disaggregate each $\ell$-node (including $\ell=0$) with fitness $x_{i_\ell}$ into any desired number of $(\ell-1)$-nodes, each with its own fitness.

This possibility allows us to perform the \emph{fine-graining} of the network, in a way that is conceptually similar to, but physically different from, the \emph{upscaling} approach in Ref.~\cite{upscaling} (which assumes a geometric embedding of nodes). 
We can therefore attach no particular meaning to the level $\ell=0$ and consider any `negative' level $m<0$ (stretching all the way down to $m=-\infty$) as well, provided that the (ultrametric) distances between all pairs of $m$-nodes are given and consistent with the higher-level ones, i.e. such that $f(d_{i_\ell,j_\ell})=f(d_{i_m,j_m})$ whenever $i_\ell=\mathbf{\Omega}_{\ell-1}\cdots\mathbf{\Omega}_m(i_m)$ and $j_\ell=\mathbf{\Omega}_{\ell-1}\cdots\mathbf{\Omega}_m(j_m)$ for all $\ell> m$.
Clearly, this requirement is always ensured in two notable cases: \emph{i)} if distances are ultrametric and the associated dendrogram is used to define which $m$-nodes branch into which $(m-1)$-nodes as we go deeper in the hierarchy of partitions; \emph{ii)} in the distance-free case $f\equiv 1$. We consider the latter an instructive example and discuss it explicitly in the rest of this section.

Note that, in general, we may start from $\ell=0$ and assign each $0$-node $i_0$ a different value of $\gamma_{i_0}$, then specify a hierarchy of coarse-grainings (and even fine-grainings) and calculate the corresponding values of $\gamma_{i_\ell}$ for all $\ell$-nodes and the resulting properties of the network, for all $\ell\ne0$. 
This leaves a lot of flexibility, in principle allowing us to taylor the resulting properties of the network to any degree of heterogeneity. 
However, to avoid making \emph{ad hoc} assumptions, we put ourselves in the simplest situation where distances are switched off (i.e. $f\equiv1$, so that the model is governed by Eq.~\eqref{eq:fitness} and is entirely non-geometric), all $0$-nodes are statistically equivalent (i.e. $\gamma_{i_0}\equiv\gamma_0$ for all $i_0=1,N_0$), and the dendrogram of coarse grainings is ${b_\ell}$-regular: at each level $\ell$, the $N_\ell$ $\ell$-nodes are merged into a number
\begin{equation}
N_{\ell+1}=\frac{N_\ell}{b_\ell}=\dots=\frac{N_0}{\prod_{m=0}^\ell b_m}
\label{eq:nodes}
\end{equation}
of $(\ell+1)$-nodes, each formed by exactly ${b_\ell}$ $\ell$-nodes.
Note that this is the most homogeneous choice, as it preserves the statistical equivalence of all the $N_\ell$ $\ell$-nodes at every hierarchical level, i.e. $\gamma_{i_\ell}\equiv\gamma_\ell$ for all $i_\ell=1,N_\ell$ where
\begin{equation} \gamma_{\ell+1}={b_{\ell}}^{1/\alpha}\gamma_{\ell}=\dots=\prod_{m=0}^\ell b_{m}^{1/\alpha}\gamma_0=\left(\frac{N_0}{N_{\ell+1}}\right)^{1/\alpha}\!\!\!\!\!\!\!\gamma_0
\label{eq:ummagumma}
\end{equation}
(with $\alpha=1/2$ here), as ensured by Eq.~\eqref{eq:gamma}.
This means that, for any $\ell$, the fitness values $\{x_{i_\ell}\}_{\ell=1}^{N_\ell}$ are i.i.d. with common distribution 
\begin{equation}
\rho_\ell(x,1/2,\gamma_\ell) =\sqrt{\frac{\gamma_\ell}{2\pi} }\frac{e^{-\gamma_\ell/(2x)}}{x^{3/2}}\quad (x>0),
\label{eq:Levy2}
\end{equation} 
effectively reducing a multivariate problem to a univariate one.
The resulting probability of generating a graph at a given hierarchical level $\ell$ does not depend on the labelling of nodes, i.e. it is unchanged upon permutations of the nodes' labels.
This property is known as \emph{exchangeability}~\cite{35,36} and is considered to be a desirable property of random graph models~\cite{pim,glee}.
In general, it is a property of all \emph{graphons}, i.e. dense limits of random graph sequences~\cite{36,graphon1,graphon2}.

\begin{figure}[t]
\includegraphics[width=\textwidth]{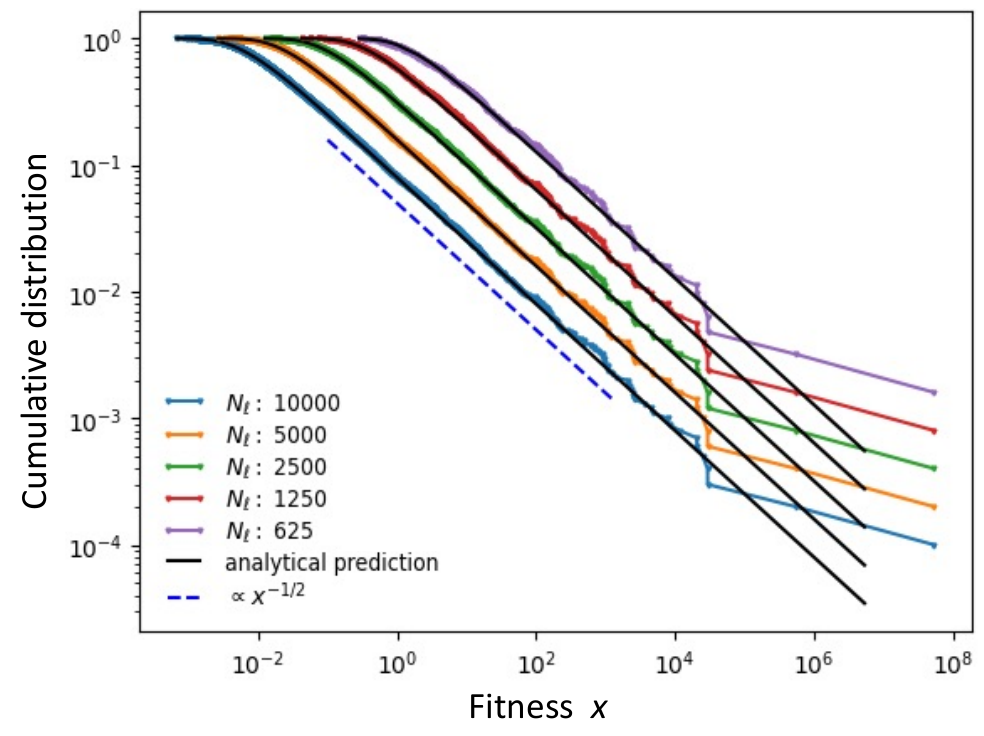}
\caption{\label{fig:levy} \textbf{Fitness distribution in the annealed scale-invariant model.} The points show the cumulative distribution of the node fitness $x$ across five different hierarchical levels ($\ell=0,1,2,3,4$), for a single realization from an $\alpha$-stable distribution with parameter choice $\alpha=1/2$, $N_0=10^4$, $b=2$. 
The solid lines are the corresponding analytical $\alpha$-stable cumulative distributions obtained integrating Eq.~\eqref{eq:Levy2} with  $\gamma_{\ell+1}={b}^{1/\alpha}\gamma_{\ell}$, $\alpha=1/2$ and $b=2$.
The dashed line is a power-law with exponent $-1/2$, confirming that the non-cumulative fitness distribution has power-law tails with exponent $-3/2$. Note that there is no upper cut-off to this tail, despite the increasing network density for higher hierarchical levels, because the fitness of a node has no bounds.
}
\end{figure}

The above prescriptions make the model similar to an annealed version of the FM~\cite{fitness} or equivalently to the class of (rank-1) IRGM~\cite{inho}, with two special requirements: \emph{i)} here the fitness is defined at all hierarchical levels simultaneously and \emph{ii)} the connection probability can only take the scale-invariant form given by Eq.~\eqref{eq:fitness}.
Note that the fitness distribution depends on the hierarchical level $\ell$ through the parameter $\gamma_\ell$, which, as clear from Eq.~\eqref{eq:ummagumma}, cannot decrease since $N_\ell$ cannot increase. This implies an overall shift towards larger values of the fitness as nodes are coarse-grained.
For instance, if we take $b_\ell=b$ (the branching ratio is level-independent), then Eq.~\eqref{eq:ummagumma} implies $\gamma_\ell=b^{\ell/\alpha}\gamma_0=b^{2\ell}\gamma_0$ and the corresponding behaviour of the fitness distribution is illustrated in Fig.~\ref{fig:levy}. Irrespective of the rightward shift, the tail of the fitness distribution is always a pure power-law proportional to $x^{-1-\alpha}$, independently of $\ell$.
We will adopt the choice $b_\ell=b$ throughout the rest of the paper, although all the  results that we obtain for a fixed hierarchical level $\ell$ hold true irrespective of this choice and are therefore general (the choice of a level-independent $b$ only affects how the calculated quantities change across hierarchical levels).

As a side remark, we note that the fine-graining procedure, if iterated indefinitely, will keep sparsifying the network until it breaks up into multiple disconnected components, and ultimately isolated nodes (\emph{dust}). In the companion paper~\cite{rajat}, we provide some rigorous results about the connectivity of the graph and the possible associated phase transitions, as a function of the model parameters.

In what follows, we are interested in characterizing the topological properties of the resulting network.
Since the fitness is annealed, the expected local topological properties involving each node $i_\ell$, when averaged over the randomness of the fitness, will be identical. However, what interests us is deriving typical structural patterns relating, irrespective of the particular realization of the fitness (and hence surving after averaging over such realizations), the correlation between different local properties of nodes. For instance, we are interested in the expected degree $k_\ell(x)$ of an $\ell$-node whose realized fitness is $x$ at level $\ell$ (note that all $\ell$-nodes with the same value $x$ of the realized fitness are statistically equivalent in the random realizations of network, so the expected degree only depends on $x$). In this way, we necessarily lose the identity of the node (since each $\ell$-node $i_\ell$ is assigned different values of the fitness $x_{i_\ell}$ in different realizations) but we keep the structural relationship between degree and fitness.
We can therefore drop the subscript $i_\ell$ accompanying any local topological property (such as $k_{i_\ell}$, $k^{nn}_{i_\ell}$, $c_{i_\ell}$) and replace it with the dependence of the expected value of that property on the realized fitness $x$.

In the rest of this section, we will use Eq.~\eqref{eq:Levy} to provide a complete analytical characterization of the annealed model when $\alpha=1/2$, although we will also retrieve similar results for all $\alpha\in(0,1)$ partially analytically (see the Appendix) and through numerical sampling of the fitness, using the suitable procedures~\cite{liang2013survey,chambers1976method,weron1995computer}. 
Further progress is possible by replacing the $\alpha$-stable PDF with a pure Pareto one with the same tail exponent $-1-\alpha$~\cite{rajat}.

\subsection{Scale-free networks from scale-invariance}
We have clarified that scale-free and scale-invariant networks are distinct concepts. 
In what follows, we show how the annealed scale-invariant model can spontaneously lead to scale-free networks, thus connecting the two concepts and providing a nontrivial recipe for generating scale-freeness purely from scale invariance.

\begin{figure}[t]
	\includegraphics[width=.98\textwidth]{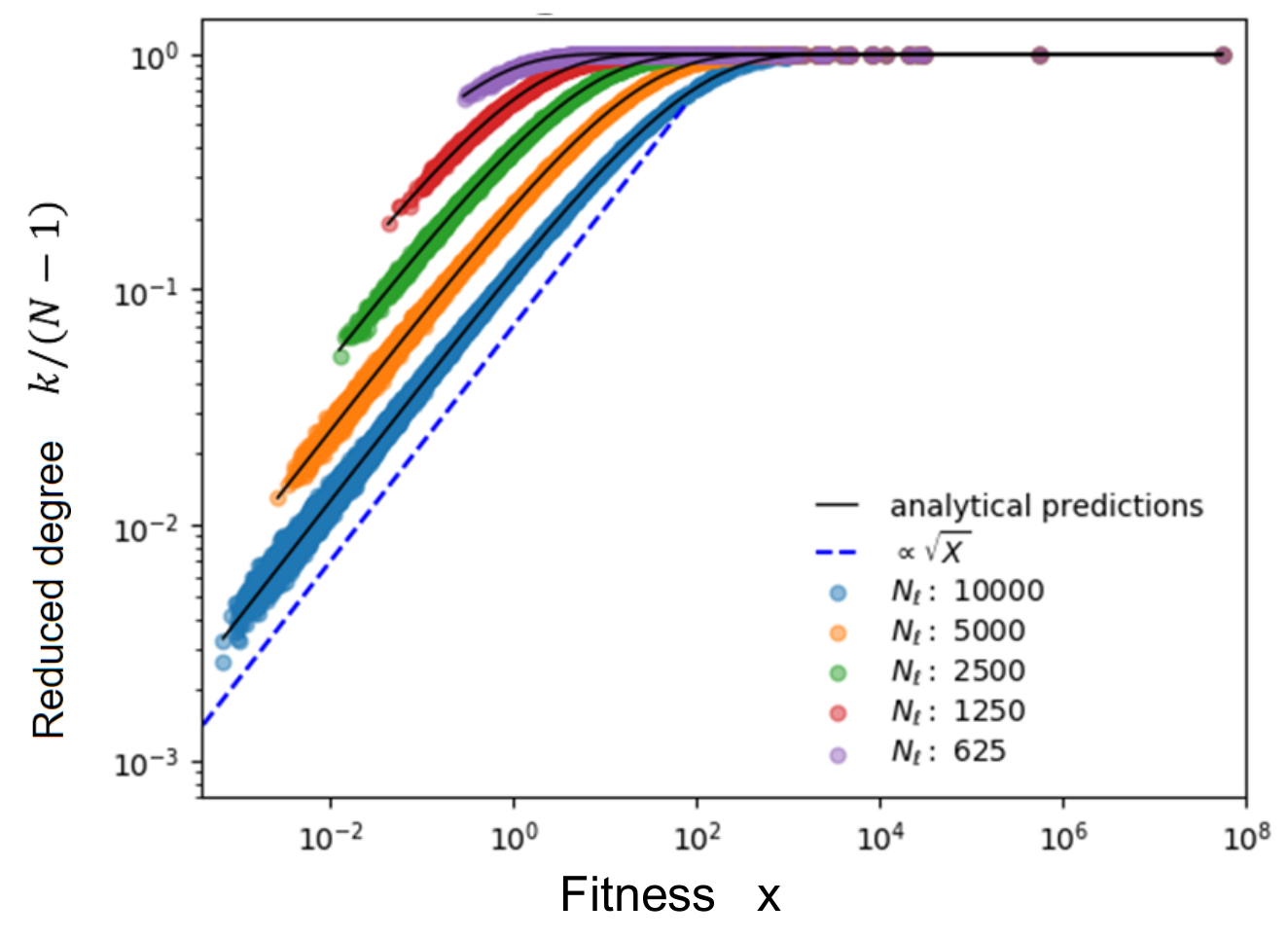}
\caption{\label{fig:kVSx} \textbf{Reduced degree as a function of fitness in the annealed scale-invariant model.} 
The circles represent the reduced degree $\kappa_\ell$ of each $\ell$-node as a function of the corresponding fitness $x$ in numerical simulations of the model across five different hierarchical levels ($\ell=0,1,2,3,4$), for the parameter choice $\alpha=1/2$, $N_0=10^4$, $b=2$. 
The solid lines are the expected theoretical relationship $\kappa_\ell(x)$ obtained via Eq.~\eqref{eq:sqrt2} for the same parameter values. The dashed line is proportional to the square root of the fitness, emphasizing the behaviour $\kappa_\ell(x) \approx{\sqrt{2\delta\gamma_\ell  x}}$ of the (reduced) degree of nodes with small fitness. For generic $\alpha\in(0,1)$, the (reduced) degree of nodes with small fitness is proportional to $x^\alpha$.}
\end{figure}

As we show in the Appendix, for $\alpha=1/2$ the expected degree $k_\ell(x)$ of an $\ell$-node with fitness $x$ is exactly calculated as
\begin{equation} \label{eq:sqrt}
k_\ell(x) =(N_\ell-1)\left(1-e^{-\sqrt{2\delta\gamma_\ell  x}}\right).
\end{equation}
It is convenient to rescale the degree $k_\ell$ by $N_\ell-1$, thereby defining the \emph{reduced degree}
\begin{equation}
\kappa_\ell\equiv\frac{k_\ell}{N_\ell-1}\in[0,1],
\label{eq:kappa}
\end{equation}
whose range is independent of $\ell$ and whose node-averaged value $\bar{\kappa}_\ell$ coincides with the network density \emph{excluding self-loops} (see the Appendix).
Clearly, Eq.~\eqref{eq:sqrt} is equivalent to 
\begin{equation}
\kappa_\ell(x) =1-e^{-\sqrt{2\delta\gamma_\ell  x}},
\label{eq:sqrt2}
\end{equation}
an exact calculation that is confirmed by numerical simulations, as shown in Fig.~\ref{fig:kVSx}. 
Note that, for $x$ sufficiently small, Eq.~\eqref{eq:sqrt} is approximated by $k_\ell(x) \approx(N_\ell-1){\sqrt{2\delta\gamma_\ell  x}}$ (or equivalently $\kappa_\ell(x) \approx{\sqrt{2\delta\gamma_\ell  x}}$), i.e. the expected degree of nodes with small fitness is proportional to the \emph{square root} of the fitness, not the fitness itself (this scaling is also confirmed in Fig.~\ref{fig:kVSx}).
For general $\alpha\in(0,1)$, it is possible to show (see the Appendix) that, if $\lambda_{\ell}(t,\alpha,\gamma_{\ell})$ denotes the LT of $\rho_\ell(x,\alpha,\gamma_\ell)$ as in Eq.~\eqref{eq:laplace}, then Eq.~\eqref{eq:sqrt2} generalizes to 
\begin{equation}
\kappa_\ell(x) =1-\lambda_{\ell}(\delta x,\alpha,\gamma_{\ell}),
\label{eq:LT}
\end{equation}
which, for $x$ sufficiently small, is approximated by $\kappa_\ell(x) \propto x^{\alpha}$.
This result beautifully illustrates the aforementioned key difference between the annealed scale-invariant model and the CM: even for very small values of the fitness, Eq.~\eqref{eq:fitness} does \emph{not} reduce to $p_{i_\ell,j_\ell}(\delta)\approx \delta x_{i_\ell} x_{j_\ell}$ and the expected degree is \emph{not} proportional to the fitness.
This is due to the divergence of all moments of the fitness in the annealed case, which implies $\max_{i_\ell}\{x_{i_\ell}\}=+\infty$ and makes the regime $\delta\ll (\max_{i_\ell}\{x_{i_\ell}\})^{-2}$ (usually assumed in the sparse CM) impossible, irrespective of the hierarchical level $\ell$.

\begin{figure}[t]
\includegraphics[width=.98\textwidth]{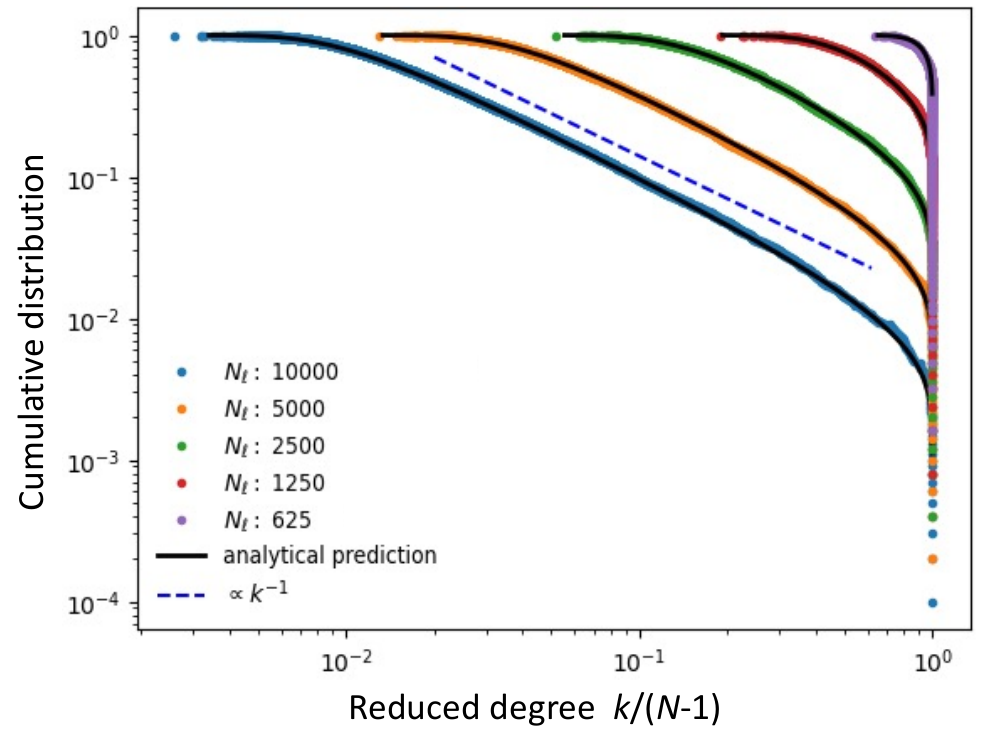}
\caption{\label{fig:Pk} \textbf{Degree distribution in the annealed scale-invariant model.} Cumulative degree distribution (fraction of nodes with reduced degree $\ge\kappa$) across five different hierarchical levels ($\ell=0,1,2,3,4$), for the parameter choice $\alpha=1/2$, $N_0=10^4$, $b=2$.  
The circles represent a single realization of the network, while the solid lines correspond to the theoretical prediction given by Eq.~\eqref{eq:kappa_distrib}.
The dashed line is a power-law with exponent $-1$, corresponding to a power-law $Q_\ell(\kappa)\propto\kappa^{-2}$ for the non-cumulative distribution. This exponent is universal for all $\alpha\in(0,1)$ and different from the exponent $-1-\alpha$ of the corresponding non-cumulative fitness distribution. Another difference is the presence of an upper cut-off $C_\ell(\kappa)$ (due to the fact that $\kappa$ cannot exceed 1) becoming stronger as the hierarchical level increases.
}
\end{figure}

As a related result, again proven in the Appendix, for $\alpha=1/2$ the expected degree distribution induced by Eqs.~\eqref{eq:fitness} and~\eqref{eq:Levy2} can be exactly calculated as
\begin{equation}
P_\ell(k)= \frac{2\gamma_\ell\sqrt{\frac{\delta}{\pi}}\exp\left[\frac{-\delta \gamma_\ell^2}{\ln^2 \left(1-\frac{k}{N_\ell-1}\right)}\right]}{(N_\ell-1-k)\ln^2 \left(1-\frac{k}{N_\ell-1}\right)}
\label{eq:degree_distrib}
\end{equation}
for $k\ge0$, and $P_\ell(k)=0$ otherwise.
The degree distribution above shows a twofold dependence on the hierarchical level $\ell$, as there are two contrasting tendencies as $\ell$ increases. On the one hand, the number of nodes $N_\ell$ decreases, hence the possible range of values $[1,N_\ell-1]$ for the degree $k$ shrinks: this implies a tendency for the degree to decrease. 
On the other hand, the ongoing coarse-graining is such that, on average, $\ell$-nodes acquire more and more links as $\ell$ increases: this implies a tendency for the degree to increase.
We can remove the effect of the first tendency by considering the probability distribution $Q_\ell(\kappa)$ for the reduced degree $\kappa_\ell$, which is easily calculated from $P_\ell(k)$ as
\begin{equation}
Q_\ell(\kappa)=\frac{P_\ell[(N_\ell-1)\kappa]}{1/(N_\ell-1)}= \frac{2\gamma_\ell\sqrt{\frac{\delta}{\pi}}\exp\left[\frac{-\delta \gamma_\ell^2}{\ln^2 \left(1-\kappa\right)}\right]}{\left(1-\kappa\right)\ln^2 \left(1-\kappa\right)}.
\label{eq:kappa_distrib}
\end{equation}
We see that the distribution has a residual dependence on the level $\ell$ through the parameter $\gamma_\ell$. 
As a consequence, the reduced degree distributions obtained for different hierarchical levels do not collapse upon each other, as confirmed in Fig.~\ref{fig:Pk} using the same parameter choice as above.
This is purely the effect of the second tendency.
Indeed we see that, as $\ell$ increases, there is a more and more pronounced accumulation of values of the reduced degree $\kappa_\ell$ close to the maximum value $1$.
This is a saturation effect cutting off the tail of the degree distribution. 

Importantly, for values of the degree that are sufficiently lower than the upper cut-off, the distribution has a universal power-law trend proportional to $\kappa^{-2}$, for all values of $\alpha\in(0,1)$ (hence without requiring a fine-tuning of $\alpha$ to a specific value in that interval). 
Indeed, one can show analytically (see the Appendix) that
the right tail of the reduced degree distribution behaves as
\begin{equation}
Q_\ell(\kappa)\approx \kappa^{-2}C_\ell(\kappa),
\end{equation}
where $C_\ell(\kappa)$ is a cut-off function with a peak at values of $\kappa$ that increase towards $1$ as $\ell$ increases. The cut-off function captures stronger and stronger finite-size effects as the network size shrinks under the effect of coarse-graining. 
In the companion paper~\cite{rajat} we identify the specific scaling for the model parameters for which the cut-off function disappears and the degree distribution is rigorously proven to have a power-law tail with universal exponent equal to $-2$, irrespective of the value of $\alpha$.
Note that, in the opposite direction (decreasing $\ell$), one can always reach the sparse regime through fine-graining, i.e. by subdividing each $\ell$-node into multiple $(\ell-1)$-nodes and so on. In such a regime, the effect of the cut-off function practically vanishes and the network is essentially scale-free with universal degree exponent $-2$. 

As anticipated, the universal exponent $-2$ for the degree distribution is different from the tail exponent $-1-\alpha\in(-2,-1)$ for the underlying fitness distribution $\rho_\ell(x)$, as a consequence of the divergence of all moments of the latter and the related non-linear dependence between degree and fitness, even for small fitness values. 
Interestingly, a mechanism producing the universal exponent $-2$ has been advocated previously~\cite{krioukov2010hyperbolic}, for instance on the basis of the fact that that exponent describes the random geometric graphs corresponding to the asymptotically de Sitter spacetime of our accelerating universe and to its large-scale Lorentzian geometry~\cite{gamma2,gamma2bis}.
More generally, the degree tail exponent $-2$ lies at the edge of the empirical range of exponents observed for the vast majority of networks, which are found in the interval $(-3,-2]$~\cite{mynatrevphys}. This empirical range of exponents is incompatible with the hypothesis that the degree itself is drawn from an $\alpha$-stable distribution with tail exponent $-1-\alpha$ (this is presumably why stable distributions have not been used in the literature to describe empirical degree distributions).  
In our model, however, the fitness does follow an $\alpha$-stable distribution as a consequence of the requirement of scale-invariance of the network, and at the same time the degree distribution has a different, realistic exponent. 
This remark further illustrates a consequence of the fact that, as we already mentioned, the degree itself cannot be renormalized exactly and in full generality, while the fitness can.

\begin{figure}[t]
	\includegraphics[width=\textwidth]{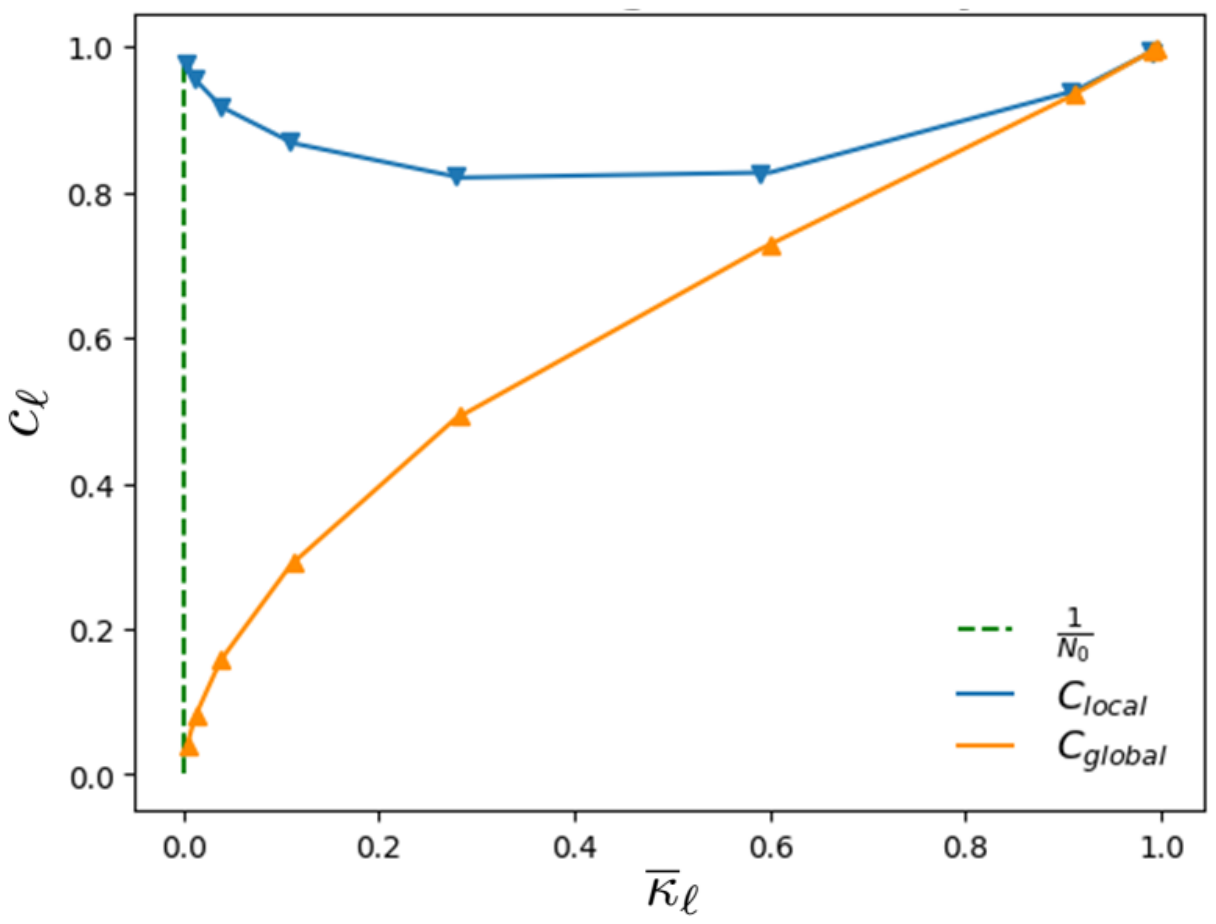}
\caption{\label{fig:CVSdens} \textbf{Local and glocal clustering coefficient as a function of density along the renormalization flow.} The average local ($c_{\ell}^\textrm{local}$) and global ($c_{\ell}^\textrm{global}$) clustering coefficients are shown as a function of the network density (excluding self-loops) $\bar{\kappa}_{\ell}$ for different coarse-grainings of the scale-invariant model with $\alpha=1/2$, $N_0=10^4$, $b=2$. Triangles refer to a single realization of the (coarse-grained) network, while the solid lines show the expected values. The dashed line is a reference corresponding to a density $1/N_0=10^{-4}$.}
\end{figure}

\begin{figure*}[t]
\includegraphics[width=\textwidth]{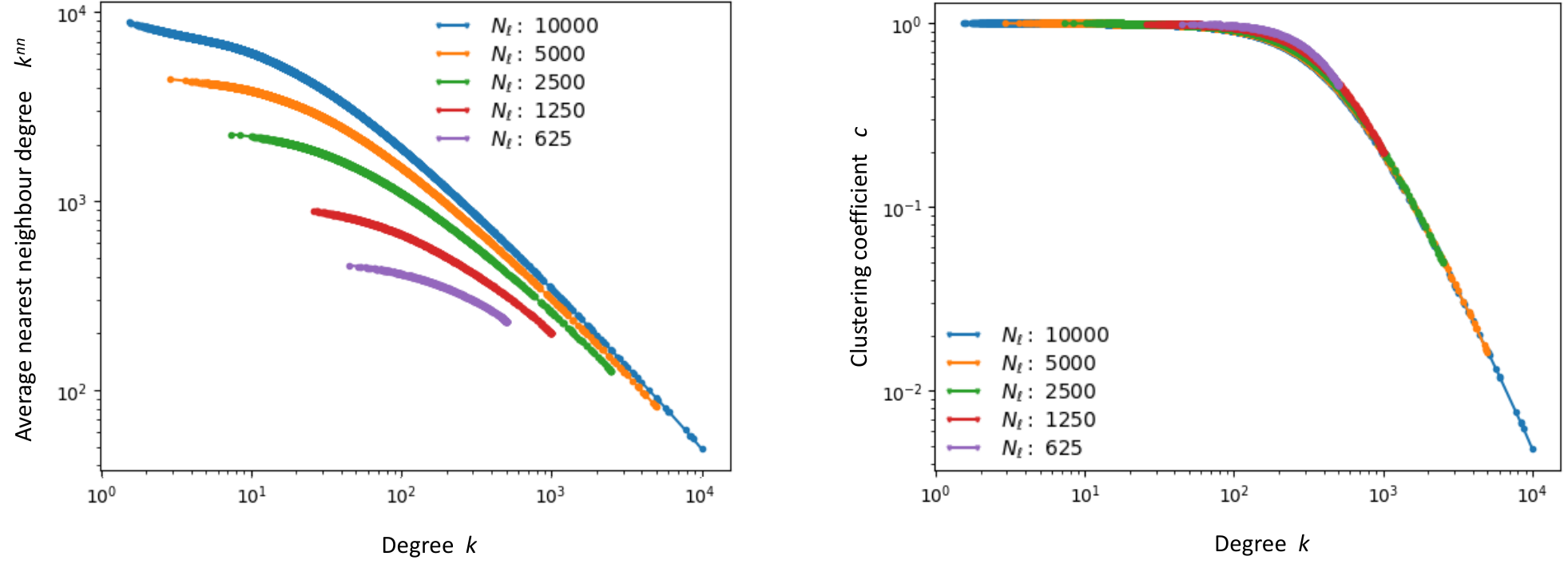}
\caption{\label{fig:knnc} \textbf{Local assortativity and clustering properties.} Average nearest neighbour degree $k^{nn}_i$ (left) and local clustering coefficient $c_i$ (right) versus degree $k_i$ in the annealed scale-invariant model across different hierarchical levels, for the parameter choice $\alpha=1/2$, $N_0=10^4$, $b=2$.}
\end{figure*}

\subsection{Assortativity and clustering without geometry}
We now show that the model leads to realistic disassortativity and clustering patterns, including a non-vanishing average local clustering coefficient even in the sparse regime.
The latter is a remarkable result, given the difficulty of generating sparse clustered networks in models with independent edges and no dependence on geometry.

In Fig.~\ref{fig:CVSdens} we show, as a function of the link density (or equivalently the average reduced degree $\bar{\kappa}_\ell$), the node-averaged local clustering coefficient $c_\ell^\textrm{local}\equiv\bar{c}_\ell=\sum_{i_\ell=1}^{N_\ell}c_{i_\ell}/N_\ell$ and the \emph{global clustering coefficient} $c_\ell^\textrm{global}\equiv \Delta_\ell/\Lambda_\ell$ obtained for different hierarchical levels.  The latter is defined as the ratio of the overall number $\Delta_\ell$ of realized triangles (each counted three times) to the number $\Lambda_\ell$ of \emph{wedges}, i.e. \emph{potential} triangles~\cite{barrat_smallworld,clust_scale-free,clust_mulini,clust_hyperbolic_glo} (see the Appendix).
Various studies have shown that, apart from cases where the network is sufficiently homogeneous~\cite{barrat_smallworld}, the average local and global clustering coefficients can be quite different~\cite{clust_scale-free,clust_mulini,clust_hyperbolic_glo,clust_hyperbolic_loc}.
In particular, an empirically widespread property of real networks is their `large' overall local clustering, defined as a nonvanishing (strictly positive) node-averaged local clustering coefficient even in the sparse regime where the network density $\bar{\kappa}_\ell$ (without self-loops), and possibly the global clustering coefficient, vanish as the number of nodes increases~\cite{local_clust}.
 
In our model, we can easily assess the behaviour of both clustering coefficients as a function of the density $\bar{\kappa}_\ell$, whose expected value can be analytically calculated for $\alpha=1/2$ (see the Appendix) as
\begin{equation}
\langle\bar{\kappa}_\ell\rangle =  1-\frac{\gamma_\ell\sqrt{\delta}}{2\pi}
\MeijerG{3}{0}{0}{3}{\cdot}{-1/2,0,0}{ {\delta\gamma_\ell^2}/{4}}
\label{eq:expdensity}
\end{equation}
where $\MeijerG{m}{n}{p}{q}{a_1,\ldots,a_p}{b_1,\ldots,b_q}{z}$ denotes the Meijer-$G$ function, which is an increasing function of the combination $\delta\gamma_\ell^2$ (therefore the network density increases as the level $\ell$ increases). 
We recall that, in the annealed scenario considered here, we are simoultaneously generating graphs at all scales $\ell=-\infty,\dots,+\infty$, ranging from the fully connected regime ($\langle \bar{\kappa}_{+\infty}\rangle=1$) to the fully disconnected one ($\langle \bar{\kappa}_{-\infty}\rangle =0$).
We can therefore inspect the expected average local clustering coefficient $c_{\ell}^\textrm{local}$ and expected global clustering coefficient $c_{\ell}^\textrm{global}$ as a function of the network density $\bar{\kappa}_\ell$. From Fig.~\ref{fig:CVSdens} we see that, remarkably, $c_{\ell}^\textrm{global}$ decreases as the density decreases (that is, as the level $\ell$ decreases), while $c_{\ell}^\textrm{local}$ retains finite values. This is in qualitative accordance with the empirical results for various real-world networks~\cite{clust_mulini}. 
In particular we find that, even for the particular hierarchical level(s) $\ell^*$ corresponding to the sparse regime $\langle \bar{\kappa}_{\ell^*}\rangle\propto 1/N_{\ell^*}$, $c_{\ell^*}^\textrm{local}$ remains finite despite the asymptotic vanishing of the network density.
In random graph models, a non-vanishing local clustering coefficient and a vanishing global clustering coefficient in the sparse regime were found also in the CM~\cite{clust_Boguna} (in the limit where the tail exponent of the degree distribution approaches the value $-2$ found here), in a class of `windmill' graphs~\cite{clust_mulini} and in the hyperbolic geometric model~\cite{clust_hyperbolic_glo,clust_hyperbolic_loc}.
In real-world networks, a typical tendency of the global clustering coefficient to be significantly smaller than the average local clustering coefficient (and even vanishing) has also been documented~\cite{clust_mulini}. 
In the companion paper~\cite{rajat}, a rigorous study of the expected value of the numbers $\Delta_\ell$ of triangles and $\Lambda_\ell$ of wedges (representing the expected value of the numerator and denominator of the global clustering coefficient respectively) is provided.

We finally consider the average nearest neighbour degree $k^{nn}_{i_\ell}$~\cite{newman_origin} and local clustering coefficient $c_{i_\ell}$~\cite{local_clust} as a function of the degree $k_{i_\ell}$ of each $\ell$-node. These quantities are plotted in Fig.~\ref{fig:knnc} for $\alpha=1/2$. The plots show decreasing trends for both $k^{nn}_{i_\ell}$ and $c_{i_\ell}$ as $k_{i_\ell}$ increases. Together with the presence of a power-law degree distribution with a cut-off, these properties are widespread in real-world networks~\cite{newman_origin,local_clust,mynatrevphys}. 
It is remarkable that all the realistic topological properties exhibited by the annealed model are generated solely from the requirement of scale-invariance.

\subsection{Erd\H{o}s-R\'enyi graphs as degenerate scale-invariant graphs\label{sec:ER}}
In retrospect, we note here that the ER model~\cite{ER} is equivalent to a particular specification of our annealed model where, at each hierarchical level $\ell$, the PDF of the fitness is a delta function $\rho_\ell(x)=\delta(x-x_\ell)$, so that all $\ell$-nodes have the same (deterministic) fitness $x_\ell$ and the connection probability has therefore the same value $p_\ell=1-\exp{(-\delta x_\ell^2)}$ for all pairs of $\ell$-nodes. The delta distribution can indeed be thought of as a degenerate stable distribution under the process that homogeneously coarse-grains the graph: when $b$ $\ell$-nodes (each with the same value $x_\ell$ of the fitness) are merged into an $(\ell+1)$-node to produce the next level $\ell+1$, all such $(\ell+1)$-nodes will have fitness equal to $x_{\ell+1}=bx_\ell$, hence still characterized by a delta-like PDF given by $\rho_{\ell+1}(x)=\delta(x-x_{\ell+1})$, and the model will remain an ER graph with renormalized connection probability $p_{\ell+1}=1-\exp{(-\delta b^2 x_\ell^2)}$. Clearly, a deterministic fitness also makes the annealed version of the model identical to the quenched version with the corresponding choice of the fitness.

In this sense, our annealed scale-invariant model and the ER model can be both interpreted as deriving from the same principle of scale-invariance under coarse-graining in homogeneous blocks (and additive fitness), the key difference being that our model allows for heterogeneous (non-deterministic) values of the fitness and therefore necessarily replaces the delta distribution with a one-sided $\alpha$-stable one. In other words, as soon as heterogeneity is introduced for the fitness, the scale-invariant requirement immediately leads from completely homogeneous ER graphs to complex networks with realistic topologies. This notable result suggests that scale-invariance and heterogeneity, taken together, might represent an effective, parsimonious mechanism for explaining several properties of real-world networks.

\section{Conclusions\label{sec:conclusions}}
We proposed a renormalization scheme based on the identification of a scale-invariant random graph model. The functional form of the probability for two nodes to be connected is independent on the hierarchical level being considered. At each level, the model can generate any network in two possible ways, with exactly the same probability: either hierarchically, by generating the finest-grained network and then coarse-graining it via progressive non-overlapping (but otherwise arbitrary) partitions, or directly, using appropriately renormalized parameters. These parameters include a global scale-invariant density parameter, a necessary set of hidden `fitness' variables attached to each (block-)node, and, if useful, a set of dyadic factors representing distances or communities. It turns out that the model possesses scale-invariance without postulating the existence of node coordinates in an underlying metric space.

If the fitness values are treated as quenched, the model can guide the renormalization of real-world graphs.
In this case, the parameters of the model can be identified with empirical quantities attached to nodes and dyads.
In our application to the ITN, we found that a one-parameter fit of the model to the observed network density is enough to accurately replicate many local topological properties of individual nodes, even across several hierarchical levels of resolution (which is related to the desirable property of projectivity~\cite{pim}). This result exemplifies the deep \emph{a priori} conceptual distinction between scale-free networks (in the sense of power-law degree distributions, which are absent in the ITN) and scale-invariant networks (in the sense of the network formation mechanisms being consistent across scales, as found in the ITN) highlighted by the model. 

If the fitness values are annealed, the model naturally leads to one-sided L\'evy-stable fitness distributions, which are characterized by a tail exponent $-1-\alpha\in(-2,-1)$.
The properties of stability and infinite divisibility of these distributions allow for the definition of a proper renormalization group in both forward (coarse-graining) and backward (fine-graining) directions. 
At the same time, the divergence of all moments of these distributions implies that the multiscale model is not asymptotically equivalent to the CM and dcSBM, showing that those models are not scale-invariant.
The annealed version of the model has also the property of exchangeability, which means that graph probabilities are unchanged upon relabelling of nodes.
It turns out that the requirement of scale invariance spontaneaously leads to scale-free networks with degree distribution featuring a universal power-law decay $P(k_\ell)\propto k^{-2}$ (which does not require a fine-tuning of $\alpha$) followed by a density-dependent cut-off and with realistic assortativity and clustering properties, without postulating mechanisms such as growth, preferential attachment or hyperbolic embedding.
In particular, in the sparse regime the model is simultaneously scale-free and locally clustered, with no need for metric distances producing clustering as a result of triangular inequalities as postulated in the hyperbolic model~\cite{krioukov2010hyperbolic}.

Importantly, the desirable topological properties generated by our annealed model differ from those of random graphs that, while being defined through the same connection probability as in Eq.~\eqref{eq:prob_renorm}, are characterized by fitness variables with finite mean~\cite{geoff,NR,CF}. Indeed, those studies did not consider a scale-invariant random graph setting and, consequently, they did not demand that in the annealed setting the fitness variables are $\alpha$-stable random variables, hence with $0<\alpha<1$ because of the positivity of the fitness. Our results indicate that many properties that are usually valid in the case $\alpha>1$ or, more generally, in the case of arbitrary fitness with finite first moment, break down in the infinite-mean regime $0<\alpha<1$. Notably, this observation suggests that the infinite-mean regime considered here is particularly important in order to capture many properties of real-world networks. Taken together, scale-invariance and heterogeneity seem to be an effective mechanism to explain those properties.

\acknowledgements
DG acknowledges support from the Dutch Econophysics Foundation (Stichting Econophysics, Leiden, the Netherlands) and the Netherlands Organisation for Scientific Research (NWO).
This work is supported by the European Union - NextGenerationEU - National Recovery and Resilience Plan (`Piano Nazionale di Ripresa e Resilienza', PNRR), project `SoBigData.it - Strengthening the Italian RI for Social Mining and Big Data Analytics' - Grant IR0000013 (n. 3264, 28/12/2021) (\url{https://pnrr.sobigdata.it/}).
This work is also supported by the project NetRes - `Network analysis of economic and financial resilience', Italian DM n. 289, 25-03-2021 (PRO3 Scuole), CUP D67G22000130001 (\url{https://netres.imtlucca.it}). 
Finally, ML and DG acknowledge support from the PAI project Pro.Co.P.E. - `Prosociality, Cognition and Peer Effects', funded by the IMT School for Advanced Studies Lucca (\url{https://procope.imtlucca.it}).

%
%
\appendix

\section{Determining the scale-invariant connection probability}
Here we show how the scale-invariance requirement stated in Eq.~\eqref{eq:Pml}, for any model with independent links as formulated in Eq.~\eqref{eq_product}, leads to the unique form of the connection probability given by Eq.~\eqref{eq:prob_renorm}.

Let us consider a partition $\mathbf{\Omega}_{\ell}$ that maps an $\ell$-graph with $N_{\ell}$ $\ell$-nodes and adjacency matrix $\mathbf{A}^{(\ell)}$ to an $(\ell+1)$-graph with $N_{\ell+1}$ $(\ell+1)$-nodes and adjacency matrix $\mathbf{A}^{(\ell+1)}$.
The relation between the entries of the matrices $\mathbf{A}^{(\ell)}$ and $\mathbf{A}^{(\ell+1)}$ is given by Eq.~\eqref{eq:Arenorm}.
Now, for any random graph model with independent links as stated in Eq.~\eqref{eq_product}, $a^{(\ell)}_{i_{\ell},j_{\ell}}$ is a Bernoulli random variable equal to $1$ with probability $p^{(\ell)}_{i_{\ell},j_{\ell}}$ and equal to $0$ with probability $1-p^{(\ell)}_{i_{\ell},j_{\ell}}$.
Similarly, $a^{(\ell+1)}_{i_{\ell+1},j_{\ell+1}}$ is a Bernoulli random variable equal to $1$ with probability $p^{(\ell+1)}_{i_{\ell+1},j_{\ell+1}}$ and equal to $0$ with probability $1-p^{(\ell+1)}_{i_{\ell+1},j_{\ell+1}}$.
Now, the scale-invariance requirement in Eq.~\eqref{eq:Pml} demands that we should create, \emph{with equal probability}, any of the possible realizations of the adjacency matrix $\mathbf{A}^{(\ell+1)}$ either by: \emph{i)} generating the possible realizations of the matrix $\mathbf{A}^{(\ell)}$ (using the associated probabilities $\{p^{(\ell)}_{i_{\ell},j_{\ell}}\}$) and then aggregating the corresponding $\ell$-graphs into $(\ell+1)$-graphs, or \emph{ii)} directly generating all the possible realizations of the matrix $\mathbf{A}^{(\ell+1)}$ (using the associated probabilities $\{p^{(\ell+1)}_{i_{\ell+1},j_{\ell+1}}\}$).
Scale-invariance also demands that $p^{(\ell)}_{i_\ell,j_\ell}$ depends on $\ell$ only through its parameters. 
Assuming that these parameters are a combination of global ($\delta_\ell$), node-specific ($x_{i_\ell}$, $x_{j_\ell}$) and dyadic ($d_{i_\ell,j_\ell}$) factors, we can write $p^{(\ell)}_{i_\ell,j_\ell}(\delta_\ell)
=p_{i_\ell,j_\ell}(\delta_\ell)$.
Enforcing scale-invariance means finding not only the functional form of $p_{i_\ell,j_\ell}$, but also the renormalization rules connecting $\delta_\ell,x_{i_\ell},x_{j_\ell},d_{i_\ell,j_\ell}$ to their next-level counterparts $\delta_{\ell+1},x_{i_{\ell+1}},x_{j_{\ell+1}},d_{i_{\ell+1},j_{\ell+1}}$.

To enforce the scale-invariance requirement, we first consider the case when the connection at the coarse-grained level $\ell+1$ involves two distinct blocks $i_{\ell+1}\ne j_{\ell+1}$. In this case, since a link between the pair $(i_{\ell+1},j_{\ell+1})$ of $(\ell+1)$-nodes is present if and only if there is at least one link present between any pair $(i_{\ell},j_{\ell})$ of $\ell$-nodes such that $i_{\ell}\in i_{\ell+1}$ and $j_{\ell}\in j_{\ell+1}$, the probability that $i_{\ell+1}$ and $j_{\ell+1}$ are \emph{not} connected is equal, according to the procedure \emph{ii)} described above, to the probability that none of the pairs of underlying $\ell$-nodes is connected. Since links are independent, this probability equals $\prod_{i_{\ell}\in i_{\ell+1}}\prod_{j_{\ell}\in j_{\ell+1}}\left[1-p_{i_{\ell},j_{\ell}}(\delta)\right]$.
On the other hand, according to the procedure \emph{i)}, the same event occurs with probability $1-p_{i_{\ell+1},j_{\ell+1}}(\delta)$. 
Enforcing the equality between the two probabilities leads to the condition
\begin{equation}
1-p_{i_{\ell+1},j_{\ell+1}}(\delta) =  \prod_{i_{\ell}\in i_{\ell+1}}\prod_{j_{\ell}\in j_{\ell+1}}\left[1-p_{i_{\ell},j_{\ell}}(\delta)\right].
\label{eq:Prenorm}
\end{equation}
Taking the logarithm of both sides of Eq.~\eqref{eq:Prenorm}, we obtain
\begin{equation}
\ln\left[1-p_{i_{\ell+1},j_{\ell+1}}(\delta)\right] =  \sum_{i_{\ell}\in i_{\ell+1}}\sum_{j_{\ell}\in j_{\ell+1}}\ln\left[1-p_{i_{\ell},j_{\ell}}(\delta)\right],
\label{eq:logPrenorm}
\end{equation}
from which we can now derive the scale-invariant form of the connection probability.
Note that Eq.~\eqref{eq:Prenorm} is consistent with taking the expected values of both sides of Eq.~\eqref{eq:Arenorm}. 
However, it cannot be derived directly in that way, because the two expected values are taken with respect to different probability distributions having different support, i.e. $P\big(\mathbf{A}^{(\ell+1)},\mathbf{\Theta}_\ell\big)$ and $P\big(\mathbf{A}^{(\ell)},\mathbf{\Theta}_\ell\big)$ respectively.
Let us first consider the case where the connection probability $p_{i_{\ell},j_{\ell}}$ does not depend on any dyadic factor $d_{i_{\ell},j_{\ell}}$. 
In this case, the only functional form of $p_{i_{\ell+1},j_{\ell+1}}$ compatible with Eq.~\eqref{eq:logPrenorm} for every pair of $(\ell+1)$-nodes is such that 
\begin{equation}
\ln\left[1-p_{i_{\ell+1},j_{\ell+1}}(\delta)\right]=-\delta\, g(x_{i_{\ell+1}})\, g(x_{j_{\ell+1}})
\label{eq:criterionlog}
\end{equation}
where $g(x)$ is a positive function such that
\begin{equation}
g(x_{i_{\ell+1}})=\sum_{i_{\ell}\in i_{\ell+1}}g(x_{i_{\ell}})
\end{equation}
and $\delta$ is positive and $\ell$-independent. Note that the positivity of $\delta$ and $g(x)$ follows from the fact that, since $0\le p_{i_{\ell+1},j_{\ell+1}}(\delta)\le1$, $\ln\left[1-p_{i_{\ell+1},j_{\ell+1}}(\delta)\right]$ has to be non-positive. On the other hand, $g(x)$ has to have the same sign for all nodes, otherwise for some pair of nodes the product $g(x_{i_{\ell+1}})\, g(x_{j_{\ell+1}})$ will be negative.
Interpreting $g(x)$ as the impact of the fitness $x$ on the connection probability, it makes sense to choose the positive sign for $g(x)$ (and, incidentally, to assume that $g(x)$ is monotonically increasing with $x$). For similar reasons, $\delta$ has to be positive as well.
Now, if the quantity $x$ is node-additive (e.g. because it is identified with some empirical additive quantity, like the GDP in our model of the ITN), then the fitness of each $(\ell+1)$-node $x_{i_{\ell+1}}$ should be consistently obtained as a sum $\sum_{i_{\ell}\in i_{\ell+1}}x_{i_{\ell}}$ over the underlying $\ell$-nodes.
This implies that, after reabsorbing any (positive) proportionality factor into $\delta$, the only possible choice for $g(x)$ in the additive case is $g(x)=x$.
By constrast, if we do not require $x$ to be node-additive, we can always invoke the desired monotonicity of $g(x)$ and redefine $x\gets g(x)$ (indeed, there is no \emph{a priori} reason why $x_{i_{\ell}}$, rather than $g(x_{i_{\ell}})$, should be regarded as the `natural' node-specific factor affecting the connection probabilities involving $i_\ell$).
This makes the redefined fitness $x$ additive by construction.
In summary, by redefining the node-specific factor $x$ in a way that makes it node-additive, and reabsorbing any global constant into $\delta$, the only possible functional form for $p_{i_\ell,j_\ell}$ under the requirement of scale-invariance (and in absence of dyadic factors) is such that
\begin{equation}
\ln\left[1-p_{i_{\ell+1},j_{\ell+1}}(\delta)\right]=-\delta\, x_{i_{\ell+1}}x_{j_{\ell+1}},
\label{eq:bilinear}
\end{equation}
for $i_{\ell+1}\ne j_{\ell+1}$, or equivalently
\begin{equation}
p_{i_\ell,j_\ell}(\delta) = 1- e^{-\delta x_{i_\ell} x_{j_\ell}},\quad \delta,x_{i_\ell},x_{j_\ell}>0,\quad i_{\ell}\ne j_{\ell},
\label{eq:prob_renormSI}
\end{equation}
where $\delta$ is scale-invariant and $x_{i_{\ell+1}}=\sum_{i_{\ell} \in i_{\ell+1}}x_{i_{\ell}}$.

Now we consider the connection probability between a block $i_{\ell+1}$ and itself, i.e. the self-loop at the coarse-grained level. In this case, to avoid double counting the internal pairs of nodes, Eq.~\eqref{eq:logPrenorm} should be replaced by the expression
\begin{equation}
\ln\left[1-p_{i_{\ell+1},i_{\ell+1}}(\delta)\right] = \!\!\! \sum_{i_{\ell}\in i_{\ell+1}}\; \sum_{j_{\ell}\in i_{\ell+1},j_{\ell} \leq i_{\ell}}\!\!\!\ln\left[1-p_{i_{\ell},j_{\ell}}(\delta)\right].
\label{eq:logPrenorm-selfloops}
\end{equation}
Now, by isolating the terms corresponding to self-loops in the quantity on the right hand side, we can rewrite the remaing terms as in Eq.~\eqref{eq:bilinear} and obtain:
\begin{equation}
\begin{split}
&\sum_{i_{\ell}\in i_{\ell+1}} \left[ \sum_{j_{\ell}\in i_{\ell+1},j_{\ell} < i_{\ell}}\!\!\!\!\!\!\ln\left[1-p_{i_{\ell},j_{\ell}}(\delta)\right] + \ln\left[1-p_{i_{\ell},i_{\ell}}(\delta)\right] \right] \\ 
&=\!\!\!\sum_{i_{\ell}\in i_{\ell+1}} \left[ \frac{1}{2}\sum_{j_{\ell}\in i_{\ell+1},j_{\ell} \neq i_{\ell}}\!\!\!\!\!\!\ln\left[1-p_{i_{\ell},j_{\ell}}(\delta)\right] + \ln\left[1-p_{i_{\ell},i_{\ell}}(\delta)\right] \right]\\ 
&=\!\!\!\sum_{i_{\ell}\in i_{\ell+1}} \left[ - \frac{\delta}{2}\sum_{j_{\ell}\in i_{\ell+1},j_{\ell} \neq i_{\ell}}  \!\!\!\!\!\!x_{i_\ell} x_{j_\ell} + \ln\left[1-p_{i_{\ell},i_{\ell}}(\delta)\right] \right].
\end{split}
\end{equation}
As argued above, the only solution for $p_{i_\ell,i_\ell}$ compatible with the requirement $x_{i_{\ell+1}} = \sum_{i_\ell \in i_{\ell+1}} x_{i_\ell} $ involves a function $\tilde{g} (x_{i_\ell})$ such that $\tilde{g}(x_{i_{\ell+1}}) = \sum_{i_\ell \in i_{\ell+1}} \tilde{g} (x_{i_\ell}) $. Now, take $\tilde{g} (x_{i_\ell}) = \sqrt{\eta} \, x_{i_\ell}$ for some $\eta > 0$. Then the requirement in Eq. \eqref{eq:logPrenorm-selfloops} finally takes the form
\begin{equation}
\eta \sum_{i_\ell\in i_{\ell+1}} \sum_{j_\ell\in i_{\ell+1}} x_{i_{\ell}}x_{j_{\ell}} = \!\!\!\sum_{i_{\ell}\in i_{\ell+1}} \left[ \frac{\delta}{2}\sum_{j_{\ell}\in i_{\ell+1},j_{\ell} \neq i_{\ell}} \!\!\!\!\!\!x_{i_\ell} x_{j_\ell} + \eta x_{i_{\ell}}^2 \right]
\label{eq:logPrenorm-selfloops2}
\end{equation}
where we have used $$\small x_{i_{\ell+1}}^2 = \left(\sum_{i_\ell \in i_{\ell+1}} x_{i_\ell}\right)^2 = \sum_{i_\ell \in i_{\ell+1}} \sum_{j_\ell\in i_{\ell+1}} x_{i_{\ell}}x_{j_{\ell}}.$$
Clearly, the only possible solution for Eq.~\eqref{eq:logPrenorm-selfloops2} is given by $\eta = \frac{\delta}{2}$, yielding:
\begin{equation}
p_{i_\ell, i_\ell} = 1 - e^{-\frac{\delta}{2} x_{i_\ell}^2},\quad \delta,x_{i_\ell},x_{j_\ell}>0.
\label{SIdiag}
\end{equation}
Taken together, Eq.~\eqref{eq:prob_renormSI} and~\eqref{SIdiag} coincide with what stated in Eq.~\eqref{eq:prob_renorm} when $f\equiv 1$, i.e. with Eq.~\eqref{eq:fitness}.

If we add dyadic factors, i.e. if we allow $p_{i_\ell,j_\ell}$ to additionally depend on some positive function $f(d)$ of the dyadic quantity $d$, while at the same time preserving the bilinear dependence of $\ln\left[1-p_{i_{\ell+1},j_{\ell+1}}(\delta)\right]$ on $x_{i_\ell}$ and $x_{j_\ell}$ (i.e. preserving the additivity of the fitness), then Eq.~\eqref{eq:bilinear} has to be generalized to 
\begin{equation}
\ln\left[1-p_{i_{\ell+1},j_{\ell+1}}(\delta)\right]=-\delta\, x_{i_{\ell+1}}\,x_{j_{\ell+1}}\,f(d_{i_{\ell+1},j_{\ell+1}})
\label{eq:mortaccistracci}
\end{equation}
for $i_{\ell+1}\ne j_{\ell+1}$ and
\begin{equation}
\ln\left[1-p_{i_{\ell+1},i_{\ell+1}}(\delta)\right]=-\frac{\delta}{2}\, x_{i_{\ell+1}}\,x_{i_{\ell+1}}\,f(d_{i_{\ell+1},i_{\ell+1}})
\label{eq:mortaccistraccistracci}
\end{equation}
otherwise, where $f(d_{i_{\ell},j_{\ell}})$ renormalizes as
\begin{equation}
x_{i_{\ell+1}}\,x_{j_{\ell+1}}\,f(d_{i_{\ell+1},j_{\ell+1}})={\sum_{i_{\ell} \in i_{\ell+1}}\sum_{j_{\ell}\in j_{\ell+1}} x_{i_{\ell}} x_{j_{\ell}} f\big(d_{i_{\ell},j_{\ell}}\big)}.
\label{eq:dyadic}
\end{equation}
Equations~\eqref{eq:mortaccistracci},~\eqref{eq:mortaccistraccistracci} and~\eqref{eq:dyadic} coincide with Eqs.~\eqref{eq:prob_renorm} and~\eqref{eq:distance}, thus completing our proof.
Note that in principle the constant $\delta$ may be entirely reabsorbed into the fitness $x$ (as mentioned above) or even into the function $f(d)$, however it is useful to keep it separate as a single parameter controlling the overall density of the graph.
Also note that if the dyadic factor $d$ is interpreted as a feature enhancing the connection probability (e.g. because it represents similarity, correlation, co-affiliation, etc.), then $f(d)$ has to be an increasing function. By contrast, if $d$ suppresses the connection probability (e.g. because it represents distance or dissimilarity), then $f(d)$ has to be a decreasing function, as in our model of the ITN.

\section{GDP, distance, and trade data}
In our analysis of the ITN, the fundamental hierarchical level $\ell=0$ is the one where each $0$-node $i_0$ corresponds to a country in the world and the fitness $x_{i_0}$ corresponds to the GDP of that country. Similarly, the distance $d_{i_0,j_0}$ corresponds to the geographic distance between the two countries ${i_0}$ and ${j_0}$ and a realized link ($a_{i_0,j_0}=1$) corresponds to the existence of a trade relation (in either direction) between ${i_0}$ and ${j_0}$.

GDP data are taken from the World Bank dataset~\cite{worldbank} and are expressed in US Dollars.
The results reported in the main body of the paper use data for year 2011. The number of countries for which GDP data are available in that year is 183. Note that, unlike the international trade data (see below), the World Bank GDP dataset covers a slightly smaller number of countries as it does not include very small ones (typically islands).

\begin{figure*}[p]
\includegraphics[width=.95\textwidth]{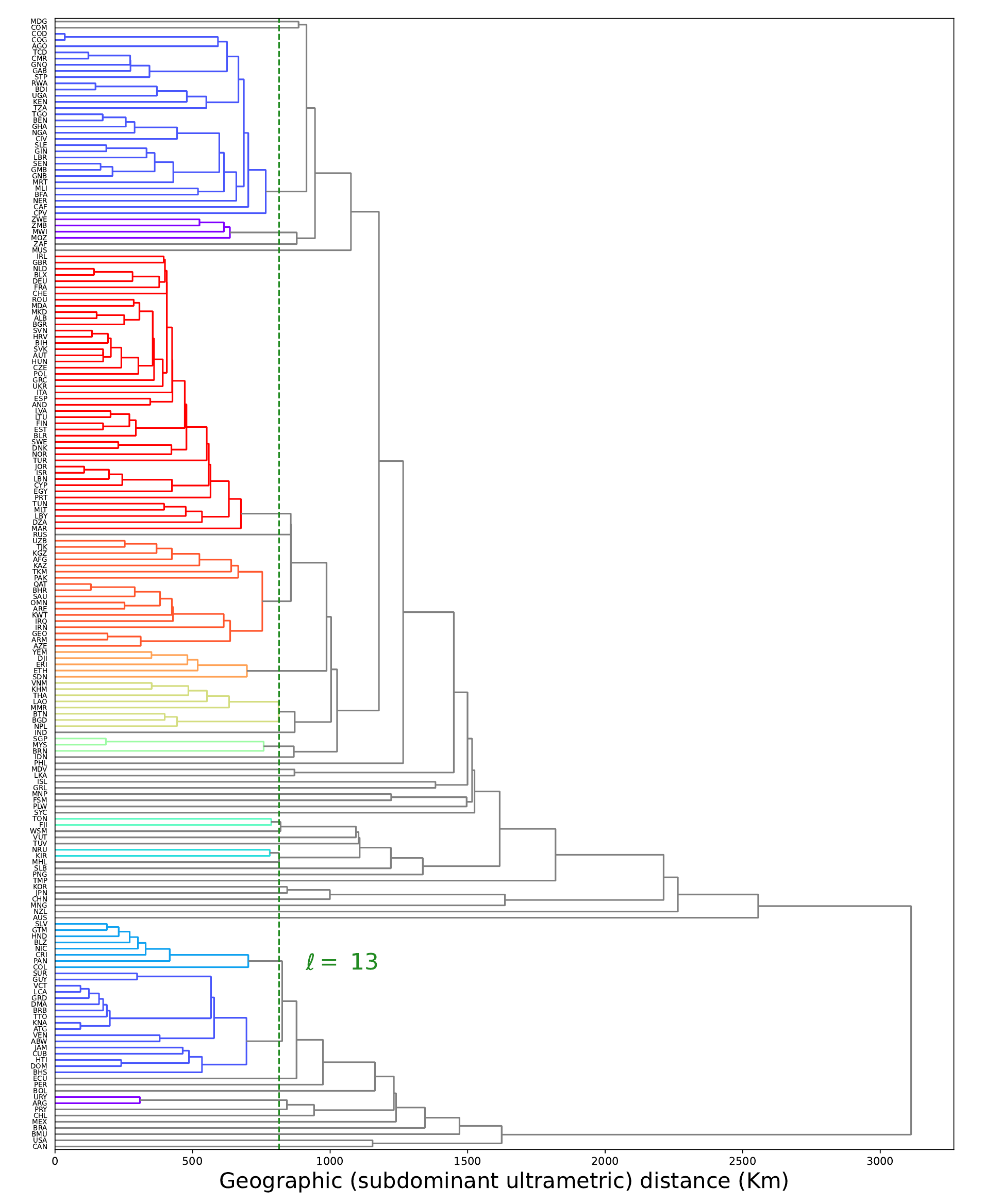}
\caption{\label{fig_Sdendro} \textbf{Dendrogram of world countries from their geographical distances using single-linkage hierarchical clustering.} The dendrogram can be used to produce any desired sequence $\{\mathbf{\Omega}_\ell\}_{\ell\ge 0}$ of geographically nested partitions, via either single-scale (straight) or multi-scale (non-straight, but monophyletic) `cuts' as explained in Fig.~\ref{fig:dendrogram}. 
In our analysis we considered 18 straight cuts at various ultrametric distances $\{h_\ell\}_{\ell=0}^{17}$ (with $h_0=0$) producing a hierarchy $\{\mathbf{\Omega}_\ell\}_{\ell=0}^{17}$ of 18 partitions and a corresponding sequence of block-countries with $N_0=183$ and $N_\ell=180-10\ell$ for $\ell=1,17$. For instance, a cut at level $\ell=13$ (dashed line) yields 50 block-countries that correspond to the 50 branches drawn in different colors.}
\end{figure*}

Geographic distance data are taken from the BACI-CEPII GeoDist database~\cite{mayer2011notes}.
It reports bilateral inter-country distances measured as population-based averages among the most populated pairs of cities across each pair of countries.
The database uses the general formula
\begin{equation}
d_{i_0,j_0}=\left(\frac{\sum_{k\in i_0}\sum_{l\in j_0}\mathrm{POP}_k\mathrm{POP}_l\,d^{\,\theta}_{k.l}}{\sum_{k\in i_0}\sum_{l\in j_0}\mathrm{POP}_k\mathrm{POP}_l}\right)^{1/\theta}
\end{equation}
developed by Head and Mayer~\cite{thetadist} for calculating the distance $d_{i_0,j_0}$ between country $i_0$ and country $j_0$ as a population-based average of the distances $d_{k,l}$ between pairs of internal agglomerations (cities, towns and places) across $i_0$ and $j_0$. The symbol $k\in i_0$ denotes that $k$ runs over the agglomerations inside country $i_0$, and $\mathrm{POP}_k$ denotes the demographic population of agglomeration $k$. In the GeoDist database, population data were taken from the World Gazetteer (\url{ https://www.world-gazetteer.com}) website.
Note that $d_{i_0,i_0}>0$, i.e. the `distance' of a country to itself is non-zero (therefore it is not a proper metric distance). This is consistent with the fact that, at higher hierarchical levels, the distance between a block-node to itself is necessarily positive as a result of the renormalization rule.
The exponent $\theta$ measures the sensitivity of trade flows to bilateral distance. As noted in the BACI-CEPII GeoDist documentation, selecting $\theta=-1$ corresponds to the usual coefficient estimated from gravity models of bilateral trade flows.
Such a choice results in the calculation of $d_{i_0,j_0}$ as a population-based average analogous to the GDP-based average used later in our own renormalization procedure when coarse-graining the network.
The agreement between our model and the ITN data actually suggests that, for the study of international trade, a better definition of inter-country distances could presumably be obtained by replacing POP with GDP in the above formula, to make inter-country distances fully consistent with our GDP-averaged renormalized distances at higher levels. Unfortunately, GDP data at the agglomeration level are much more difficult to obtain than the corresponding population data. 
For this reason, we used population-averaged distances in our analysis at level $\ell=0$, and their GDP-averaged renormalized values at higher levels $\ell>0$. Given the pairwise geographical distances $\{d_{i_0,j_0}\}_{i_0,j_0=1}^{N_0}$ at level $\ell=0$, we constructed the dendrogram of nested partitions $\{\mathbf{\Omega}_\ell\}_{\ell\ge 0}$ of world countries (shown in Fig.~\ref{fig_Sdendro}) using single-linkage hierarchical clustering, which produces \emph{subdominant ultrametric distances} $\{d^<_{i_0,j_0}\}_{i_0,j_0=1}^{N_0}$ as explained. 
A straight cut in the dendrogram induces a hierarchical level $\ell$ and a corresponding partition of countries into $\ell$-countries. The renormalized GDPs and distances are then calculated using Eqs.~\eqref{eq:GDP} and~\eqref{eq:geodist} (using the original distances).

For the construction of the International Trade Network, we used the BACI-Comtrade dataset \cite{gaulier2010baci}. 
The dataset reports the international trade flows between 207 countries for the years 2008 to 2011. 
From the full set of countries, we selected the 183 countries for which we could find matching GDP data in the World Bank database (as explained above).
In the BACI-Comtrade dataset, trade is disaggregated into 96 commodity classes labeled at a 2-digit resolution level and is expressed in thousands of dollars. The database is the result of an adjustment procedure \cite{gaulier2010baci} which reconciles unbalanced trade values as reported by importers and exporters. 
For the purpose of this study, we first merged the disaggregated data into a unique aggregate undirected network, where the monetary flows between countries is the total trade (both import and export) in all the 96 commodities, and then considered its binary (i.e. unweighted) projection. Therefore, a binary link in the $0$-graph of the ITN is present if the two countries at its endpoints have a positive trade (either import or export) in any commodity, consistently with similar analyses of the topology of the ITN constructed from different datasets~\cite{mywtw,mydouble,myassaf1,myassaf2}.
This procedure defines the empirical adjacency matrix $\tilde{\mathbf{A}}^{(0)}$ of the $0$-graph of the ITN. The empirical matrices $\tilde{\mathbf{A}}^{(\ell)}$ for $\ell>0$ are obtained via coarse-graining the empirical $0$-graph (following the general procedure illustrated in Fig.~\ref{fig:coarse_graining}) using the nested partitions $\{\mathbf{\Omega}_\ell\}_{\ell\ge 0}$ induced by the dendrogram in Fig.~\ref{fig_Sdendro}.

\section{Network properties: empirical and expected values}
Here we define the key topological properties considered in our analysis and modelling of the ITN.
Each such property is a function $Y(\mathbf{A}^{(\ell)})$ of the $N_\ell\times N_\ell$ adjacency matrix $\mathbf{A}^{(\ell)}$ (with entries $a^{(\ell)}_{i_\ell,j_\ell}=0,1$) of the generic $\ell$-graph. 
Note that this matrix is symmetric and can contain non-zero entries along the diagonal, representing self-loops. These self-loops may or may not be present in the $0$-graph, but are in any case eventually generated by the coarse graining procedure if the nodes mapped onto the same block-node are connected among themselves. 
When analysing the ITN, the relevant matrix $\mathbf{A}^{(\ell)}$ is the empirical matrix $\tilde{\mathbf{A}}^{(\ell)}$ obtained at the hierarchical level $\ell$ from the BACI-Comtrade data in year 2011 as described above. The corresponding empirical value of each topological property $Y$ of interest will be denoted as $\tilde{Y}\equiv Y(\tilde{\mathbf{A}}^{(\ell)})$.
When considering the multiscale model, $\mathbf{A}^{(\ell)}$ is instead a random (symmetric) matrix whose entries $\{a^{(\ell)}_{i_\ell,j_\ell}\}$ are Bernoulli random variables with expected value
\begin{eqnarray}
\langle a^{(\ell)}_{i_\ell,j_\ell}\rangle&=&p_{i_\ell,j_\ell}(\delta)\\
&=& \begin{cases}
 1- e^{-\delta\, \mathrm{GDP}_{i_\ell} \mathrm{GDP}_{j_\ell} /d_{i_\ell,j_\ell}} &\textrm{if}\quad i_\ell \neq j_\ell\label{eq:langle}\\
  1- e^{-\frac{\delta}{2}\,\mathrm{GDP}^2_{i_\ell}/d_{i_\ell,i_\ell}} & \textrm{if}\quad i_\ell = j_\ell
\end{cases}\nonumber
\end{eqnarray} 
where, consistently with the possible presence of self-loops, we allow for $i_\ell=j_\ell$.
Equation~\eqref{eq:langle} allows us to calculate the expected value of several topological properties.
For instance, the total number of $\ell$-links (including possible self-loops) at level $\ell$ is given by
\begin{equation}
L_\ell(\mathbf{A}^{(\ell)})=\sum_{i_\ell=1}^{N_\ell}\sum_{j_\ell=1}^{i_\ell}a^{(\ell)}_{i_\ell,j_\ell}.
\label{eq:Ll}
\end{equation}

Before considering other properties, we note that we fix the only free parameter $\delta$ to the unique value $\tilde{\delta}$ such that the expected number 
\begin{equation}
\langle L_0\rangle=\sum_{i_0=1}^{N_0}\sum_{j_0=1}^{i_0}p_{i_0,j_0}({\delta})
\label{eq:L0exp}
\end{equation}
of links of the $0$-graph equals the empirical value
\begin{equation}
\tilde{L}_0=L_0(\tilde{\mathbf{A}}^{(0)})=\sum_{i_0=1}^{N_0}\sum_{j_0=1}^{i_0}\tilde{a}^{(0)}_{i_0,j_0}=12018
\label{eq:L0tilde}
\end{equation}
observed in the ITN in year 2011. This selects the value
$\tilde{\delta} = 3.6 \cdot 10^{-17} (\mathrm{USD})^{-2}$, where $\mathrm{USD}$ stands for US dollars (the unit of measure used in GDP data).
Having fixed $\tilde{\delta}$, we can generate unbiased realisations $\{\mathbf{A}^{(\ell)}\}$ of the $\ell$-graphs from the multiscale model at any desired hierarchical level $\ell$ by sampling $\ell$-links independently with probability $\tilde{p}_{i_\ell,j_\ell}\equiv p_{i_\ell,j_\ell}(\tilde{\delta})$.
By averaging the value $Y(\mathbf{A}^{(\ell)})$ of any topological property of interest over such realizations, we can efficiently estimate the corresponding expected value 
\begin{equation}
\langle{Y}\rangle\equiv\sum_{\mathbf{A}^{(\ell)}\in\mathcal{G}_{N_\ell}}P\big(\mathbf{A}^{(\ell)},\tilde{\delta}\big)Y(\mathbf{A}^{(\ell)}),
\end{equation}
where $P\big(\mathbf{A}^{(\ell)},{\delta}\big)$ is given by Eq.~\eqref{eq_product2}, without actually calculating the above sum explicitly.
If $Y(\mathbf{A}^{(\ell)})$ is linear in $\mathbf{A}^{(\ell)}$, we can even calculate $\langle{Y}\rangle$ exactly by directly replacing $a^{(\ell)}_{i_\ell,j_\ell}$ with $\tilde{p}_{i_\ell,j_\ell}$ in the definition of $Y(\mathbf{A}^{(\ell)})$, without sampling any graph at all.
This is indeed the case for the number of links in Eq.~\eqref{eq:Ll}.

Given any $\ell$-graph $\mathbf{A}^{(\ell)}$ (be it the empirical $\ell$-graph or a random realization from the model), the main topological properties of interest to us are: the \emph{link density} 
\begin{equation}
D_\ell(\mathbf{A}^{(\ell)})\equiv\frac{2L_\ell(\mathbf{A}^{(\ell)})}{N_\ell(N_\ell+1)}=\frac{2\sum_{i_\ell=1}^{N_\ell}\sum_{j_\ell=1}^{i_\ell}a^{(\ell)}_{i_\ell,j_\ell}}{N_\ell(N_\ell+1)}
\label{eq:density}
\end{equation}
(representing the ratio of realized to maximum number of links, including possible self-loops),
the \emph{degree}  
\begin{equation}
k_{i_\ell}(\mathbf{A}^{(\ell)})\equiv\sum_{j_\ell\ne i_\ell}a^{(\ell)}_{i_\ell,j_\ell}
\label{eq:degree}
\end{equation}
(counting the number of links of the $\ell$-node $i_\ell$, excluding self-loops),
the \emph{rescaled degree}  
\begin{equation}
\kappa_{i_\ell}(\mathbf{A}^{(\ell)})\equiv\frac{1}{N_\ell-1}\sum_{j_\ell\ne i_\ell}a^{(\ell)}_{i_\ell,j_\ell}
\label{eq:rescaleddegree}
\end{equation}
(which ranges in $[0,1]$, irrespective of the vertex and hierarchical level considered), the \emph{average nearest neighbour degree}~\cite{newman_origin} 
\begin{equation}
k^{nn}_{i_\ell}(\mathbf{A}^{(\ell)})\equiv\frac{\sum_{j_\ell\ne i_\ell}\sum_{k_\ell\ne j_\ell}a^{(\ell)}_{i_\ell,j_\ell}a^{(\ell)}_{j_\ell,k_\ell}}{\sum_{j_\ell\ne i_\ell}a^{(\ell)}_{i_\ell,j_\ell}}
\label{eq:annd}
\end{equation}
(representing the average degree of the neighbours of $i_\ell$), and finally the local clustering coefficient~\cite{local_clust}
\begin{equation}
c_{i_\ell}(\mathbf{A}^{(\ell)})\equiv\frac{\sum_{j_\ell\ne i_\ell}\sum_{k_\ell\ne i_\ell,j_\ell}a^{(\ell)}_{i_\ell,j_\ell}a^{(\ell)}_{j_\ell,k_\ell}a^{(\ell)}_{k_\ell,i_\ell}}{\sum_{j_\ell\ne i_\ell}\sum_{k_\ell\ne i_\ell,j_\ell}a^{(\ell)}_{i_\ell,j_\ell}a^{(\ell)}_{k_\ell,i_\ell}}
\label{eq:clustering}
\end{equation}
(representing the number of triangles into which $i_\ell$ partipates, divided by the maximum realizable number of triangles, given the value of $k_{i_\ell}$).
All the above quantities can be averaged over nodes to obtain the following overall properties:
\begin{eqnarray}
\bar{k}_\ell(\mathbf{A}^{(\ell)})&\equiv&\frac{1}{N_\ell}\sum_{i_\ell=1}^{N_\ell}k_{i_\ell}(\mathbf{A}^{(\ell)}),\label{eq:kbar}\\
\bar{\kappa}_\ell(\mathbf{A}^{(\ell)})&\equiv&\frac{1}{N_\ell}\sum_{i_\ell=1}^{N_\ell}\kappa_{i_\ell}(\mathbf{A}^{(\ell)}),\label{eq:kappabar}\\
\bar{k}^{nn}_\ell(\mathbf{A}^{(\ell)})&\equiv&\frac{1}{N_\ell}\sum_{i_\ell=1}^{N_\ell}k^{nn}_{i_\ell}(\mathbf{A}^{(\ell)}),\label{eq:knnbar}\\
\bar{c}_\ell(\mathbf{A}^{(\ell)})&\equiv&\frac{1}{N_\ell}\sum_{i_\ell=1}^{N_\ell}c_{i_\ell}(\mathbf{A}^{(\ell)}).\label{eq:cbar}
\end{eqnarray}
Note that $\bar{\kappa}_\ell(\mathbf{A}^{(\ell)})\in[0,1]$ in Eq.~\eqref{eq:kappabar} coincides with the link density \emph{excluding self-loops}, representative an alternative to the definition of density in Eq.~\eqref{eq:density} (where self-loops are included).
Besides the average local clustering coefficient $\bar{c}_\ell(\mathbf{A}^{(\ell)})$, it is possible to define the \emph{global clustering coefficient}~\cite{clust_scale-free,clust_mulini,clust_hyperbolic_glo}
\begin{equation}
c_\ell^\textrm{global}(\mathbf{A}^{(\ell)})\equiv \frac{\Delta_\ell(\mathbf{A}^{(\ell)})}{\Lambda_\ell(\mathbf{A}^{(\ell)})}
\end{equation}
where 
\begin{equation}
\Delta_\ell(\mathbf{A}^{(\ell)})\equiv{\sum_{i_\ell=1}^{N_\ell}\sum_{j_\ell\ne i_\ell}\sum_{k_\ell\ne i_\ell,j_\ell}a^{(\ell)}_{i_\ell,j_\ell}a^{(\ell)}_{j_\ell,k_\ell}a^{(\ell)}_{k_\ell,i_\ell}}
\end{equation}
is the overall number of \emph{realized} (`closed') triangles and
\begin{equation}
\Lambda_\ell(\mathbf{A}^{(\ell)})\equiv{\sum_{i_\ell=1}^{N_\ell}\sum_{j_\ell\ne i_\ell}\sum_{k_\ell\ne i_\ell,j_\ell}a^{(\ell)}_{i_\ell,j_\ell}a^{(\ell)}_{k_\ell,i_\ell}}
\end{equation}
is the number of ($\Lambda$-shaped) \emph{wedges}, i.e. \emph{potential} (both `open' and `closed') triangles (note that each realized triangle is counted three times by both $\Delta_\ell$ and $\Lambda_\ell$).

It is important to stress that, of all the quantities defined in Eqs.~\eqref{eq:density}-\eqref{eq:cbar} for each $\ell$-node ($i_\ell=1,N_\ell$) and/or all levels ($\ell=0,17$), only the overall density ${D}_0$ of the $0$-graph is replicated by construction via the parameter choice $\delta=\tilde{\delta}$: indeed, having enforced $\langle{L}_0\rangle=\tilde{L}_0$ by equating Eqs.~\eqref{eq:L0exp} and~\eqref{eq:L0tilde} coincides with having required $\langle{D_0}\rangle=\tilde{D}_0$.
For all the other properties, including $D_\ell$ for all $\ell>0$, the agreement between the model and the empirical network is highly nontrivial and hence remarkable.

\section{Analytical form of the degree distribution for $\alpha=1/2$}
Here we derive the functional form of the expected degree distribution in the annealed model with L\'evy-distributed fitness (i.e. $\alpha=1/2$) as specified in Eq.~\eqref{eq:Levy2} and distance-independent connection probability  (i.e. $f\equiv 1$) as given by Eq.~\eqref{eq:fitness}.
To this end, for any fixed hierarchical level $\ell$ we adapt the procedure outlined in Ref.~\cite{fitness} to compute, for a typical realization of the fitness values, the distribution $P_\ell(k)$ of expected (over the realizations of the network) degrees from the PDF of the fitness $\rho_\ell(x)$ and the connection probability $p_{i_\ell,j_\ell}$, written as a function $p_{i_\ell,j_\ell}=f(x_{i_\ell},x_{j_\ell})$ of the fitness of the nodes involved, where in our case
\begin{equation}
f(x,y)=1-e^{-\delta\, x\,y}.
\end{equation}

We first notice that, since $f(x,y)$ is an increasing function of both its arguments, the expected degree $\langle k_{i_\ell}\rangle$ 
\begin{equation}
\langle k_{i_\ell}\rangle =\sum_{j_\ell\ne i_\ell}p_{i_\ell,j_\ell}=\sum_{j_\ell\ne i_\ell}f(x_{i_\ell},x_{j_\ell})
\label{eq:expdegree}
\end{equation}
is an increasing function of the fitness $x_{i_\ell}$. 
Indeed, any two $\ell$-nodes with the same fitness have the same expected degree, and $\ell$-nodes with higher fitness have larger expected degree.
For a large number $N_\ell$ of $\ell$-nodes, the above discrete sum can be approximated by an integral over the number $(N_\ell-1)\rho_\ell(y,\alpha,\gamma_\ell)$ of $\ell$-nodes (except $i_\ell$ itself) with fitness in a neighbourhood of $y$: if $k_\ell(x)$ denotes the expected degree of a node with fitness $x$ at level $\ell$, we have
\begin{eqnarray} 
k_\ell(x) &=&(N_\ell-1)\int_0^{\infty}  f(x,y)   \rho_\ell(y,\alpha,\gamma_\ell)  \mathrm{d}y\nonumber \\ 
&=&(N_\ell-1)\left(1-\int_0^{\infty}  e^{-\delta x y}  \rho_\ell(y,\alpha,\gamma_\ell)  \mathrm{d}y\right)\nonumber \\ 
&=&(N_\ell-1)\left(1-\lambda_{\ell}(\delta x,\alpha,\gamma_{\ell})\right)\label{tumadre}
\end{eqnarray} 
where $\lambda_{\ell}(t,\alpha,\gamma_{\ell})$ denotes the LT of $\rho_\ell(x,\alpha,\gamma_\ell)$ as in Eq.~\eqref{eq:laplace}.

When $\alpha=1/2$, the LT can be calculated explicitly as
\begin{eqnarray}
\lambda_{\ell}(\delta x,1/2,\gamma_{\ell})&=&\int_{0}^{\infty}e^{-\delta x y}\sqrt{\frac{\gamma_\ell}{2\pi} }\frac{e^{-\gamma_\ell/(2y)}}{y^{3/2}} \mathrm{d}y\nonumber  \\
&=&e^{-\sqrt{2\delta\gamma_\ell  x}},
\end{eqnarray}
so that
\begin{equation}
k_\ell(x)=(N_\ell-1)\left(1-e^{-\sqrt{2\delta\gamma_\ell  x}}\right),
\label{eq:kVSx}
\end{equation}
proving Eq.~\eqref{eq:sqrt}.
Inverting, we find that the fitness $x_\ell$ of an $\ell$-node with expected degree $k$ at level $\ell$ is
\begin{equation}
x_\ell(k)=\frac{1}{2 \delta\gamma_\ell }\ln^2\left(\frac{N_\ell-1}{N_\ell-1-k}\right),
\label{eq:xk}
\end{equation}
which implies
\begin{equation}
\frac{\mathrm{d}}{\mathrm{d}k}x_\ell(k)= \frac{\ln\left(\frac{N_\ell-1}{N_\ell-1-k}\right)}{\delta\gamma_\ell(N_\ell-1-k)}.
\label{eq:dxk}
\end{equation}

We can use the above expressions in order to obtain the distribution $P_\ell(k)$ of the expected degrees from the distribution $\rho_\ell(x,\alpha,\gamma_\ell)$ of the corresponding fitness. 
Indeed, starting from the fundamental equation
\begin{equation}
P_\ell(k)\mathrm{d}k=\rho_\ell\big(x_\ell(k),\alpha,\gamma_\ell\big)\mathrm{d}x_\ell(k)
\label{eq:funda}
\end{equation}
relating the probability distributions of the two random variables $k$ and $x$, and using Eqs.~\eqref{eq:Levy2},~\eqref{eq:xk} and~\eqref{eq:dxk}, we arrive at the explicit form of the distribution of expected degrees:
\begin{eqnarray}
P_\ell(k) &=& \rho_\ell\big(x_\ell(k),1/2,\gamma_\ell\big)\frac{\mathrm{d}}{\mathrm{d}k}x_\ell(k)\nonumber\\
& =& \frac{2\sqrt{\frac{\delta\gamma_\ell^2}{\pi}}\exp\left[\frac{-\delta \gamma_\ell^2}{\ln^2 \left(\frac{N_\ell-1}{N_\ell-1-k}\right)}\right]}{(N_\ell-1-k)\ln^2 \left(\frac{N_\ell-1}{N_\ell-1-k}\right)}
\end{eqnarray}
for $k\ge0$, and $P_\ell(k)=0$ otherwise.
This proves Eq.~\eqref{eq:degree_distrib}.

We can obtain the expected link density $\langle \bar{\kappa}_\ell\rangle$ (excluding self-loops) as
\begin{eqnarray}
\langle \bar{\kappa}_\ell\rangle&=& \frac{1}{N_\ell-1} \int_0^{N_\ell-1} P_\ell(k) k\, \mathrm{d}k\label{eq:2L} \\
&=& \left(1- \frac{2}{\pi}\int_0^\infty e^{-t_\ell^2-{\gamma_\ell \sqrt{\delta}}/{t_\ell}}  \mathrm{d}t_\ell \right),\nonumber
\end{eqnarray} 
where we have changed variables by introducing
\begin{equation}
t_\ell = \frac{\gamma_\ell \sqrt{\delta}}{ \ln{\frac{N_\ell-1}{N_\ell-1-k}}}.
\end{equation}
The integral in Eq.~\eqref{eq:2L} can be expressed in terms of one of the Meijer-$G$ functions $\MeijerG{m}{n}{p}{q}{a_1,\ldots,a_p}{b_1,\ldots,b_q}{z}$.
The resulting expected density can be written as 
\begin{equation}
\langle \bar{\kappa}_\ell\rangle = 1-\frac{\gamma_\ell\sqrt{\delta}}{2\pi}\MeijerG{3}{0}{0}{3}{\cdot}{-1/2,0,0}{ {\delta\gamma_\ell^2}/{4}},
\end{equation}
proving Eq.~\eqref{eq:expdensity}.

\section{The scale-free range with universal inverse square exponent} 
\begin{figure}
	\includegraphics[width=\textwidth]{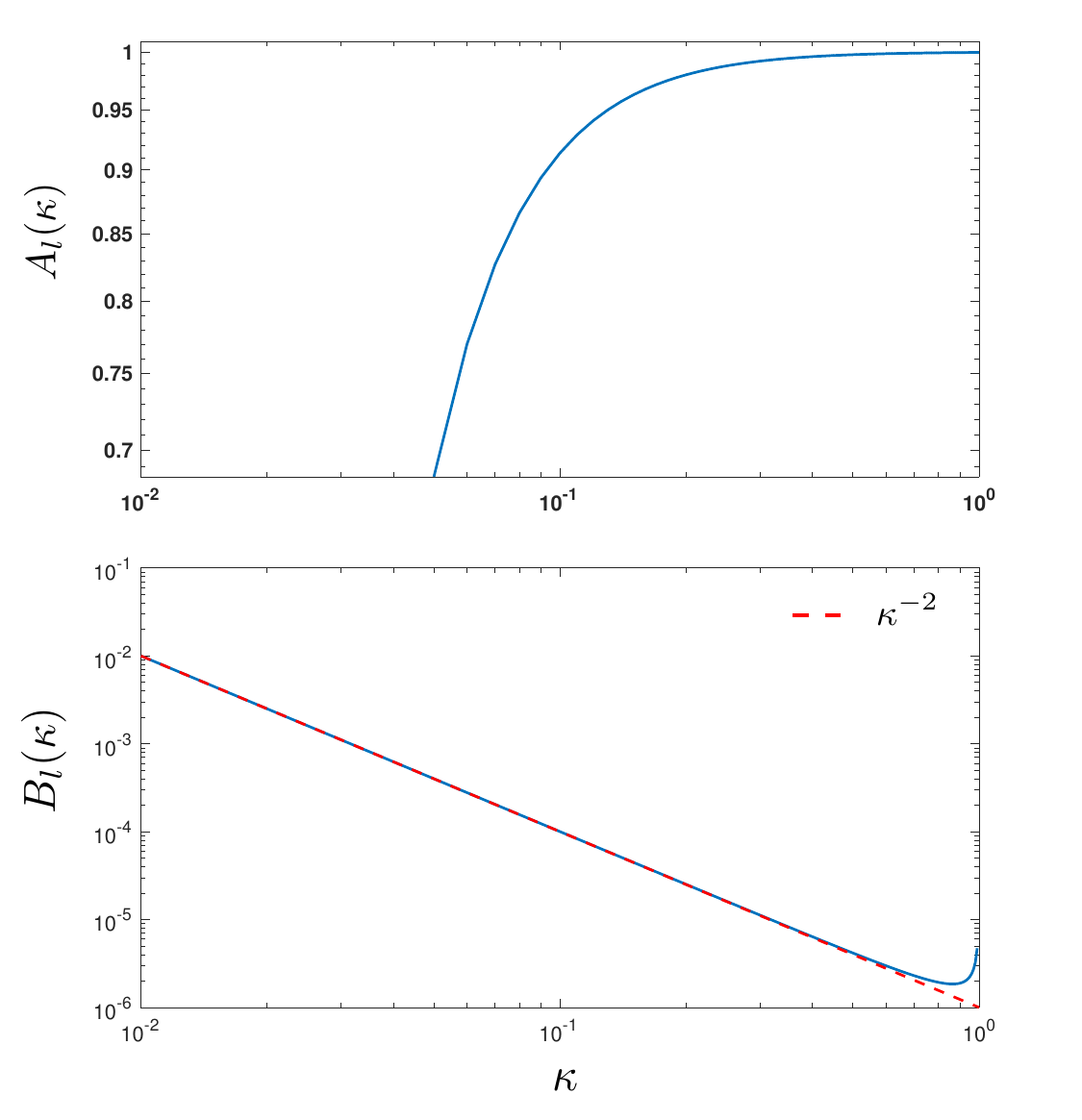}
\caption{\label{fig:Q} \textbf{The two factors contributing to the cumulative distribution of the rescaled degree.} Top: lower cut-off function $A_\ell(\kappa)$ defined in Eq.~\eqref{eq:A}. The function rapidly saturates to $A_\ell(\kappa)\approx 1$	as the rescaled degree $\kappa$ increases. 
Bottom: tail function $B_\ell(\kappa)$ defined in Eq.~\eqref{eq:B}. The function behaves as a power law $B_\ell(\kappa)\approx\kappa^{-2}$ (red dashed line) for a wide range of $\kappa$ and has an $\ell$-dependent upper cut-off corresponding to nodes whose rescaled degree saturates to 1.}
\end{figure}

We first consider the case $\alpha=1/2$ and rewrite the distribution of the reduced degree $\kappa$ shown in Eq.~\eqref{eq:kappa_distrib} as
\begin{equation}
Q_\ell(\kappa)=A_\ell(\kappa)B_\ell(\kappa)
\end{equation}
where
\begin{eqnarray}
A_\ell(\kappa)&\equiv&\exp\left[\frac{-\delta \gamma_\ell^2}{\ln^2 \left(1-\kappa\right)}\right],\label{eq:A}\\
B_\ell(\kappa)&\equiv&\frac{2\sqrt{\delta\gamma_\ell^2/\pi}}{\left(1-\kappa\right)\ln^2 \left(1-\kappa\right)}.
\label{eq:B}
\end{eqnarray}
The term $A_\ell(\kappa)$ is a lower cut-off that rapidly saturates to one as $\kappa$ increases (see Fig.~\ref{fig:Q}). 
On the other hand, $B_\ell(\kappa)$ has an intermediate power-law regime (for values of $\kappa$ not too close to 1) and an upper cut-off (for $\kappa$ closer to 1). 
This behaviour can be understood by using the expansion of $\ln(1-y) = -\sum_{n=1}^\infty (- y)^n/n=y+R(y)$ for $|y|< 1$, where $R(y)\equiv-\sum_{n=2}^\infty (- y)^n/n$: 
\begin{eqnarray}
B_\ell(\kappa)  &=& \frac{2\gamma_\ell\sqrt{\delta/\pi}}{(1-\kappa)\ln^2 (1-\kappa)}\nonumber\\
&=& \frac{2\gamma_\ell\sqrt{\delta/\pi}}{(1-\kappa) [\kappa+R(\kappa)]^2}\nonumber\\
&\approx&\left\{\begin{array}{ll}
		\frac{2\gamma_\ell\sqrt{\delta/\pi}}{\kappa^2}&\kappa\ll 1\\
+\infty&\kappa \to 1^-\end{array}\right.\nonumber\\
		&\approx&\kappa^{-2}C_\ell(\kappa)\label{taccitua}
\end{eqnarray}
where $C_\ell(\kappa)$ is a cut-off function being equal to $2\gamma_\ell\sqrt{\delta/\pi}$ for $\kappa\ll 1$ and diverging when $\kappa \to 1^-$. This is confirmed in Fig.~\ref{fig:Q}. 
Putting the pieces together, the right tail of the reduced degree distribution behaves as
\begin{equation}
Q_\ell(\kappa)\approx \kappa^{-2}C_\ell(\kappa)
\label{eq:mosoccazzi}
\end{equation}
where $C_\ell(\kappa)$ is the cut-off function. This proves our statement and is confirmed by the numerical simulations in Fig.~\ref{fig:Pk}.

Now we can partly extend the above results to the general case $\alpha\in(0,1)$ using the following argument (an alternative derivation is provided in Ref.~\cite{rajat}). 
We note from Eq.~\eqref{tumadre} that, for any $\alpha\in(0,1)$, the expected degree is uniquely determined by the LT of the fitness distribution. Even if the explicit form of $\rho_\ell(x,\alpha,\gamma_\ell)$ is not known for $\alpha\ne 1/2$ (apart from expressions involving integral representations~\cite{onesided3,onesided4,onesided5}), the LT is known and given by Eq.~\eqref{eq:laplace}. Using that formula, thereby selecting without loss of generality the value $\gamma_\alpha\equiv[\textrm{cos}(\alpha\pi/2)]^{1/\alpha}$, we see that Eq.~\eqref{tumadre} can be rewritten as
\begin{eqnarray} 
k_\ell(x) &=&(N_\ell-1)\left(1-\lambda_{\ell}(\delta x,\alpha,\gamma_\alpha)\right)\nonumber\\
&=&(N_\ell-1)\left(1-e^{-{(\delta x)}^\alpha}\right).\label{tupadre}
\end{eqnarray} 
Indeed, for $\alpha=1/2$ and $\gamma_{1/2}=[\textrm{cos}(\pi/4)]^{2}=1/2$, the above equation reduces exactly to Eq.~\eqref{eq:kVSx}.
In complete analogy with the case $\alpha=1/2$, Eq.~\eqref{tupadre} implies that, for small values of $x$, the expected degree behaves as 
\begin{equation}
k_\ell(x)\propto x^\alpha \qquad (x\ll\delta^{-1}),
\label{tusorella}
\end{equation}
while for large values of $x$ there is a saturation $k_\ell(x)\approx N_\ell-1$ (as in Fig.~\ref{fig:kVSx}) which produces the cut-off in the degree distribution $P_\ell(k)$.
Therefore, in order to establish the behaviour of $P_\ell(k)$ before the cut-off appears (i.e. for $k\ll N_\ell-1$), it is enough to invert Eq.~\eqref{tusorella} as $x_\ell(k)\propto k^{1/\alpha}$ and use it into Eq.~\eqref{eq:funda} to obtain
\begin{eqnarray}
P_\ell(k) &=& \rho_\ell\big(x_\ell(k),\alpha,\gamma_\alpha\big)\frac{\mathrm{d}}{\mathrm{d}k}x_\ell(k)\nonumber\\
& \propto& \big(x_\ell(k)\big)^{-1-\alpha}k^{-1+1/\alpha}\nonumber\\
& \propto& k^{-1-1/\alpha}k^{-1+1/\alpha}\nonumber\\
& \propto& k^{-2}\qquad\qquad\qquad\qquad\quad (k\ll N_\ell-1)\qquad
\end{eqnarray} 
where we have used $\rho_\ell(x,\alpha,\gamma_\alpha)\propto x^{-1-\alpha}$ for large enough $x$. As clear from Eq.~\eqref{tusorella}, the range of values of $x$ for which both $x_\ell(k)\propto k^{1/\alpha}$ and $\rho_\ell(x,\alpha,\gamma_\alpha)\propto x^{-1-\alpha}$ are valid is larger when $\delta$ is smaller (correspondingly, the effect of the cut-off in the degree distribution is weaker). So for sparser networks the regime $P_\ell(k)\propto k^{-2}$ is valid for a larger fraction of the range of values of $k$.
Correspondingly, the reduced degree distribution behaves as
\begin{equation}
Q_\ell(\kappa)\propto \kappa^{-2}\qquad (\kappa\ll1)
\label{tufratello}
\end{equation}
and is followed by an upper cut-off for $\kappa\lesssim1$. 
For sparser networks, Eq.~\eqref{tufratello} is valid for a larger range of values.
The above results confirm Eq.~\eqref{eq:mosoccazzi}, which was obtained for $\alpha=1/2$, and extend it to the entire range $\alpha\in(0,1)$.
In the companion paper~\cite{rajat}, the universality of the tail exponent $-2$ of the degree distribution is rigorously confirmed by replacing the $\alpha$-stable PDF of the fitness with a pure Pareto distribution with the same tail exponent $-1-\alpha$.


\clearpage
\end{document}